\pdfoutput=1
\documentclass[prd,amsmath,amsfonts,floatfix,nofootinbib,preprintnumbers]{revtex4}

\usepackage{graphicx}
\usepackage{slashed}
\usepackage{color}
\usepackage{amsmath, amssymb}
\usepackage{amssymb}

\def\la{\langle}
\def\ra{\rangle}

\def\beq{\begin{equation}}
\def\eeq{\end{equation}}
\def\bea{\begin{eqnarray}}
\def\eea{\end{eqnarray}}
\def\barr{\begin{array}}
\def\earr{\end{array}}

\def\exponential{{\mathrm{e}}}

\begin{document}

\begin{titlepage}


\vskip 0.5cm

\begin{center}
{\Large \bf The Yang-Mills gradient flow and SU(3) gauge theory\\with 12 massless
      fundamental fermions in a colour-twisted box} 
\vskip1cm {\large\bf 
C.-J.~David~Lin$^{a}$,
Kenji~Ogawa$^{a}$,
Alberto~Ramos$^{b}$
}\\ \vspace{.5cm}
{\normalsize {\sl 
$^{a}$ Institute of Physics, National Chiao-Tung University, Hsinchu 30010, Taiwan\\
$^{b}$ PH-TH, CERN, CH-1121 Geneva 23, Switzerland
}}

\vskip1.0cm {\large\bf Abstract:\\[10pt]} \parbox[t]{\textwidth}{{
We perform the step-scaling investigation of the running coupling
constant, using the gradient-flow scheme,  in SU(3) gauge theory with twelve
massless fermions in the fundamental representation. 
The Wilson plaquette gauge
action and massless unimproved staggered fermions are used in the simulations.   
Our lattice data are prepared at high accuracy, such that the statistical
error for the renormalised coupling, $g_{_{{\rm GF}}}$, is at the
subpercentage level.  
To investigate the reliability of the continuum extrapolation,
we employ two different lattice
discretisations to obtain $g_{_{{\rm GF}}}$.  
For our simulation setting, the
corresponding gauge-field averaging radius in the gradient flow has to be almost half of
the lattice size, in order to have this extrapolation under control.
We can determine the renormalisation
group evolution of the
coupling up to $g^{2}_{_{{\rm GF}}} \sim 6$, before the onset of the bulk
phase structure.  In this infrared regime, the running of the
coupling is significantly
slower than the two-loop perturbative prediction, although we cannot
draw definite conclusion regarding possible infrared conformality of this theory.
Furthermore, we comment on the issue regarding the continuum
extrapolation near an infrared fixed point.  In addition to adopting the
fit ans\"{a}tz {\it a'la} Symanzik for performing this task, we
discuss a possible alternative procedure inspired by
properties derived from low-energy scale invariance at strong coupling.
Based on this procedure, we propose a finite-size
scaling method for the renormalised coupling as a means to search for
infrared fixed point.
Using this method, it can be shown that
the behaviour of the theory around $g^{2}_{_{{\rm GF}}} \sim 6$ is
still not governed by possible infrared conformality.}}
\end{center}
\end{titlepage}

\section{Introduction}
\label{sec:introduction}
The discovery of a Higgs-like light scalar state at the Large Hadron
Collider (LHC) has stimulated a significant amount of studies in
various electroweak symmetry breaking (EWSB) models which can accommodate such
a state.  The Standard Model (SM) contains a fundamental
scalar Higgs field, and is successful in explaining all the
experimental results hitherto.  Nevertheless, the scalar sector
of the SM is widely believed to be trivial~\cite{Aizenman:1981zz,
  Frohlich:1982tw, Luscher:1988gc}, therefore the cut-off is indispensable.   This
makes its predictions of low-energy quantities, such as the Higgs
boson mass, sensitive to new physics effects which can appear as
higher-dimensional irrelevant operators~\cite{Branchina:2013jra,
  Gies:2013fua, Chu:2015nha, Akerlund:2015fya}.    In view of this, it is desirable to
find an alternative EWSB model in which the interaction is described by a relevant operator.
There are various approaches to achieve this while having a light
scalar state in the spectrum.   
Amongst these, one possibility is that this scalar particle
is a dilaton~\cite{Elander:2009pk, Elander:2012fk, Matsuzaki:2012mk, Evans:2015qaa} in the composite Higgs, or walking technicolour (WTC), 
scenario~\cite{Holdom:1984sk, Yamawaki:1985zg, Appelquist:1986an}.  In this
scenario, it is necessary to construct a gauge theory which exhibits
asymptotic freedom and quasi scale invariance in the infrared (IR)
regime.   The Goldstone boson (the dilaton) resulting from the breaking of the IR scale
invariance can be parametrically light compared to all the other
states~\cite{Dietrich:2005jn, Appelquist:2010gy}.
In addition to the possible existence of a light scalar state,  WTC models also contain other appealing
features. 
Any such model can incorporate the generation of fermion masses and the
origin of flavours, withal dynamical suppression of flavour-changing
neutral current (FCNC) processes.

%
%

The search for gauge theories that are viable for WTC
model building has been a popular subject in the lattice
community~\cite{Lucini:2015noa,DeGrand:2015zxa}\footnote{Besides the
  lattice approach, one popular technique for the search of
  WTC models is the gravity/gauge duality.  There has
  also been many works on this topic.  See Ref.~\cite{Piai:2010ma} for
an introduction to the subject.}.   This subject has led to studies of
phase structure in lattice gauge theories with many flavours of
fermions~\cite{Damgaard:1997ut, Cheng:2011ic, Miura:2011mc, deForcrand:2012vh, Huang:2014xwa},
broadening our understanding of the lattice regularisation.
In addition to numerical
calculations, it is also resulting in progress in analytic work
related to
lattice gauge theory~\cite{DelDebbio:2013sta, DelDebbio:2013qta, DelDebbio:2013zaa,
  DelDebbio:2010jy, DelDebbio:2010ze, Nogradi:2012dj, Suzuki:2013gza}.     One 
important task in this research avenue is to determine the conformal
window, 
\beq
  N_{f}^{\rm cr} \le N_{f} < N_{f}^{\rm AF}, 
\label{eq:conformal_window}
\eeq
for a gauge
theory coupled to $N_{f}$ fermions in a particular
representation\footnote{One can also consider a gauge theory containing fermions belonging to
  more than one representation~\cite{Marciano:1980zf}, and $N_{f}^{{\rm
      cr}}$ is in general lower for higher-dimensional
  representations~\cite{Sannino:2004qp, Dietrich:2006cm}.}, where 
$N_{f}^{\rm  AF}$ is the number of fermions above which asymptotic freedom is lost,
and $N_{f}^{\rm cr}$ is the ``critical'' number of fermions below
which the theory is confining in the IR.  A candidate theory for the
WTC scenario is believed to have $N_{f}$ just below $N_{f}^{\rm cr}$.
This makes the determination of $N_{f}^{\rm cr}$ an endeavour with
phenomenological importance.

%
%

Amongst the intensive investigations of candidate WTC theories using the lattice
technique, the study of SU(3) gauge theory with twelve flavours
of fermion in the fundamental representation has a long history.
Several groups of authors found that this theory is conformal in the
IR~\cite{Cheng:2014jba,Aoki:2012eq,Appelquist:2007hu,Appelquist:2009ty,
  Appelquist:2011dp,Appelquist:2012sm,Itou:2012qn,Cheng:2013eu,DeGrand:2011cu,Deuzeman:2009mh,Hasenfratz:2011xn,Miura:2011mc,Ishikawa:2013tua}.
Nevertheless, in 
Refs.~\cite{Fodor:2009wk,Fodor:2011tu} it was argued that chiral symmetry
is broken in this theory.   Amidst all these works,  one popular
approach is the step-scaling method for computing the running coupling
constant, which was originally formulated in the Schr\"{o}dinger-functional (SF) scheme~\cite{Luscher:1992zx,Heller:1997pn}.   
This method was first used to
determine the low-energy behaviour of SU(3) gauge theory with $N_{f}=12$ in
Ref.~\cite{Appelquist:2007hu}, where the authors claimed the existence
of an infrared fixed point (IRFP) by studying the coupling constant in the
SF scheme.  
Recently, we adopted the same procedure in a different
renormalisation scheme, namely the 
Twisted Polyakov Loop (TPL)
scheme~\cite{Luscher:1985wf,deDivitiis:1993hj,deDivitiis:1994yp}, 
in our investigation of the same theory, and found evidence for IR
conformality as well~\cite{Lin:2012iw}\footnote{The approach used in 
  the study of SU(3) gauge theory with twelve flavours in
  Ref.\cite{Cheng:2014jba} is similar to the step-scaling
  method.}\footnote{The author of Ref.~\cite{Itou:2012qn} used almost
  identical lattice data with a slightly different analysis procedure to
  reach the same conclusion.  In the rest of this article, we will use
our own previous work, Ref.~\cite{Lin:2012iw}, to illustrate the features of the
TPL-scheme coupling.}.  

\begin{figure}[t]
\begin{center}
\hspace{-0.5cm}
\includegraphics[scale=0.6]{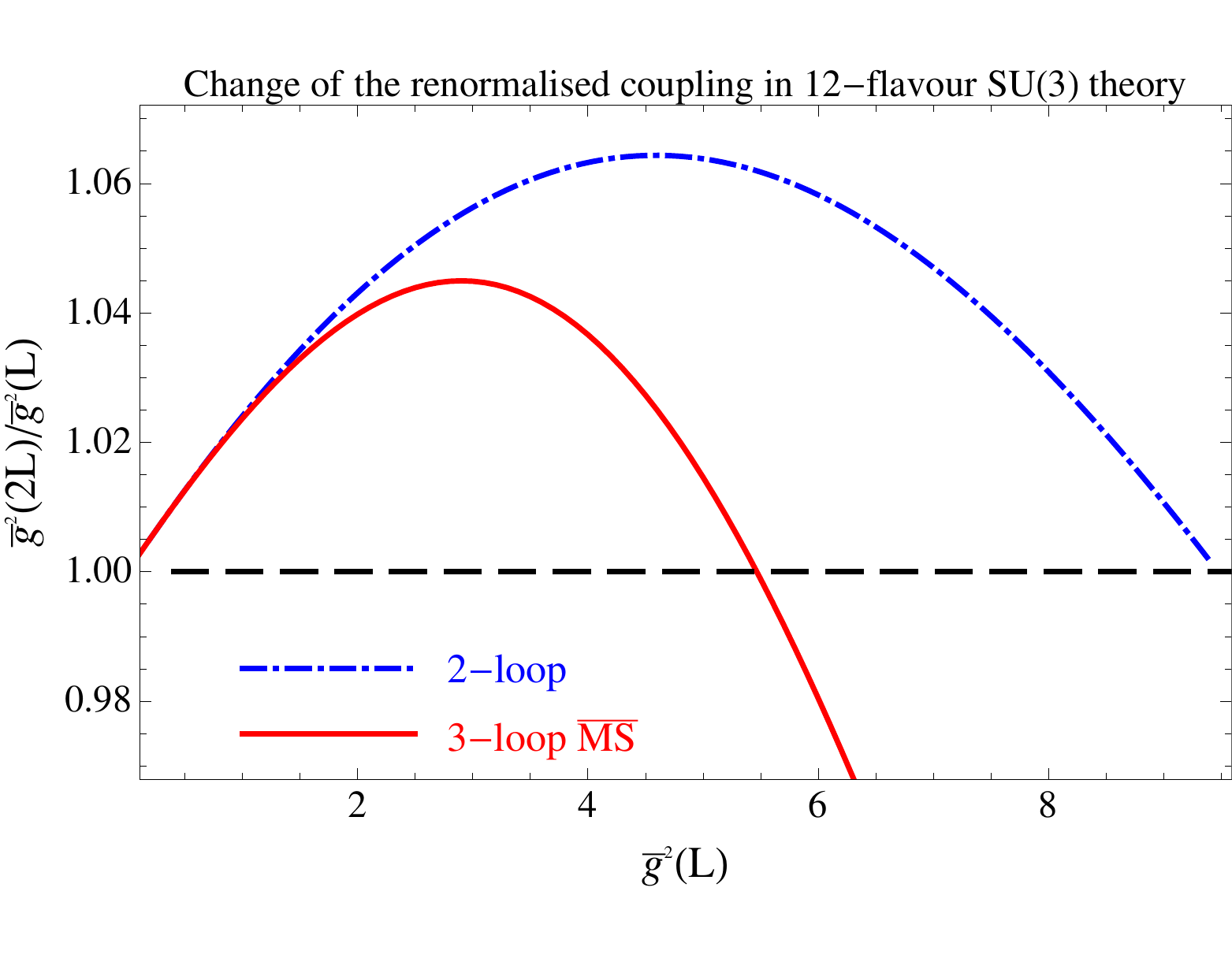}
\caption{Change of the renormalised coupling with respect to varying
  the length scale by a factor of two in SU(3) gauge theory with 12
  massless fermions in the fundamental representation.  We use the symbol $\bar{g}(L)$ to denote a generic coupling renormalised at the length scale $L$. The blue dash-dot curve is the two-loop result, while the red solid curve is from the three-loop $\overline{{\rm MS}}$-scheme computation.  This plot shows that the $\beta{-}$function is very small in this theory, and perturbative calculations predict the existence of and IRFP at strong coupling.}
\label{fig:PT_sigma}
\end{center}
\end{figure}

With the lattice computation for the search of IR-conformal field
theories becoming mature, the importance of controlling errors in
such calculations is now receiving growing attention.  To illustrate
this point as relevant to our work presented here, in
Fig.~\ref{fig:PT_sigma} we plot the perturbative predictions for the
change of a generic renormalised coupling constant with respect to the
doubling of the length scale in 12-flavours SU(3) gauge
theory\footnote{We thank Anna Hasenfratz for sending us the three-loop 
$\overline{{\rm MS}}$-scheme result used in this figure.}.   This plot
indicates that the $\beta{-}$function is very small in this theory,
and there can be an IRFP at strong coupling where perturbation theory
may not be applicable.   To obtain concrete
evidence for confirming or ruling out the existence of this IRFP, one
can resort to the lattice numerical implementation of the step-scaling 
method.  However, Figure~\ref{fig:PT_sigma} also shows that in order to
make any statistically-meaningful statement in such lattice studies,
one would have to control the error to the sub-percentage level.
Without this accuracy, it is difficult to demonstrate that the theory
flows out of the vicinity of the asymptotically-free ultraviolet fixed
point (UVFP), and then flows towards the IRFP when increasing the length
scale.  In previous step-scaling calculations for SU(3) gauge theory
with twelve flavours, such statistical
precision was not achieved.  In addition, it was difficult to estimate
the systematic effects in the continuum extrapolation in those
calculations.  This makes it challenging to draw reliable conclusions
from these computations.  For instance, in our previous step-scaling
work employing the TPL scheme~\cite{Lin:2012iw}, the lattice
simulations were performed to give statistical error around $2\%$
in the extracted coupling constant\footnote{It is well known that the
  TPL-scheme coupling is very noisy.  In Ref.~\cite{Lin:2012iw}, it is
necessary to have more than one million Hybrid Monte-Carlo
trajectories to achieve $2\%$ statistical error for some of the data
points.  This makes it challenging to further improve the error.}.  The evidence for IR scale
invariance discovered there was subsequently shown to be unreliable,
after the addition of data at a larger volume that enables better
estimation of the continuum-extrapolation error~\cite{Ogawa:Latt2013}. 

In this project,  we perform lattice simulations with massless unimproved
staggered fermions and the Wilson
plaquette gauge action from which the Yang-Mills gradient
flow~\cite{Luscher:2010iy} is implemented.
As pointed out in
Ref.~\cite{Fodor:2012td},  this method allows for the 
extraction of the renormalised coupling in the ``gradient-flow (GF)
scheme'',  {\it via} computing the energy
density of the Yang-Mills field.    The step-scaling approach is
carried out with the step-size $s=2$, on the lattice sizes
\beq
 \hat{L} = L/a = (8, 10, 12)  \longrightarrow 
   2 \hat{L} = (16, 20, 24) ,
\label{eq:step_scaling_summary}
\eeq
where $L$ is dimensionfull, and $a$ is the lattice spacing.
Since our procedure only involves the 
calculation of the smallest Wilson loops on the lattice, we can determine the 
renormalised coupling with high statistical accuracy.     Using about
one-hundred thousand  Hybrid Monte-Carlo (HMC) trajectories, we are
able to control the statistical errors to be within $0.5\%$ for the
renormalised couplings
computed on our lattice data.  Furthermore, we adopt two different 
discretisations, namely the plaquette and the clover operators, in
obtaining the energy density.   This enables us to investigate the
reliability of our results in the continuum limit.  It has been well known
that the continuum extrapolation is the main source of systematic
error in the step-scaling study, therefore this feature of our
analysis is welcome.

The current article is a report for our analysis of the GF-scheme coupling
constant in SU(3) gauge theory with twelve fermions in the fundamental
representation.  
In this work, we are able to probe the
low-energy regime of this theory, with the bare coupling $g_{0}^{2} \sim 1.45$ and the
corresponding renormalised coupling $g^{2}_{{\rm GF}} \sim 6$,  using lattices as large as
$\hat{L}=24$.   In this regime, the coupling runs significantly slower than the
two-loop perturbative prediction.  However, our work does not allow us
to draw definite conclusion regarding the existence of an IRFP in this
theory.   
At bare couplings larger than 1.45, we begin to
observe the onset of the bulk phase structure of the lattice theory.
This means that the investigation of the theory at stronger
renormalised coupling can only be
achieved with simulations at larger lattice volume, such as
$\hat{L}=32$.  This is beyond the scope of this project.

In this article, we discuss in detail the application of the
continuum extrapolation ans\"{a}tz {\it a'la} Symanzik in lattice
studies for the conformal windows in gauge theories.   We point out
that one has to be cautious in using this conventional method for
confirming IR scale invariance.  When the theory is tuned to be
close to the possible IRFP, the continuum extrapolation may have to be
conducted using a formula containing an unknown power of the lattice
spacing.   Since this kind of extrapolation is very challenging to
carry out in practice, we propose a finite-size scaling method based
on the same IR scaling property.  
We perform this finite-size scaling test in this work.  The result of this test indicates
that at $g^{2}_{{\rm GF}} \sim 6$, the behaviour of theory is not governed by
IR conformality.

%
%

This paper is organised in the following way.  In
Sec.~\ref{sec:strategy_and_simulation} we discuss our strategy and
lattice simulations.  We then present the details of our analysis and results in
Sec.~\ref{sec:analysis}.   Section~\ref{sec:FSS} contains our comment
on the continuum extrapolation, and the proposal of 
a finite-size scaling test of the renormalised coupling in the
strong-coupling regime.   
We compare our result with previous lattice computations in 
Sec.~\ref{sec:comparison}, and conclude in Sec.~\ref{sec:conclusion}.

\section{Strategy and the lattice simulation}
\label{sec:strategy_and_simulation}
\subsection{The colour-twisted boundary condition}
\label{sec:TBC}
In this work, we make use of
twisted boundary condition (TBC)~\cite{'tHooft:1979uj}, where 
the gauge field is periodic up to a gauge transformation ($\mu, \nu =
1, 2, 3, 4$ are the Lorentz indices)
\begin{equation}
  A_\mu(x+\hat \nu L_\nu) = \Omega_\nu(x)A_\mu(x)\Omega_\nu(x)^\dagger + 
  \Omega_\nu(x)\partial_\mu\Omega_\nu(x)^\dagger ,
\end{equation}
where $L_{\nu}$ is the linear size in the $\nu$ direction
(with $\hat{\nu}$ being the unit vector). The $SU(3)$ matrices
$\Omega_\mu(x)$ are called twist matrices and must obey the
consistency relation
\begin{equation}
  \Omega_\mu(x+L_\nu\hat \nu)\Omega_\nu(x) = e^{2\pi\imath n_{\mu\nu}/N}
  \Omega_\nu(x+L_\mu\hat \mu)\Omega_\mu(x)\,,
\end{equation}
Where $n_{\mu\nu}$ is an anti-symmetric tensor of integers modulo
3 called twist tensor. The concrete choice of twist matrices is
largely irrelevant, since, they change under gauge
transformations. All the physical information about the twisted
boundary conditions is contained in the twist tensor $n_{\mu\nu}$. Our
particular choice consists in using 
\begin{equation}
  n_{\mu\nu} = - n_{\nu\mu} = \left\{ 
    \begin{array}{ll}
      1& \mu = 1 \text{ and } \nu = 2 \\
      0 & \text{otherwise}
    \end{array}
  \right.
\end{equation}

On the lattice this particular choice can be realised by choosing
the twist matrices constant in space and obeying the relations
\bea
&& \Omega_{1} \Omega_{2}
= \exponential^{i2\pi/3} \Omega_{2} \Omega_{1} ,
\nonumber \\
\label{eq:omega_12_relations}
&&
\Omega_{3} = \Omega_{4} = {\mathbf{1}} .
\eea
A concrete representation is 
\beq
\label{eq:omega_12_explicit}
\Omega_{1} = \left ( \begin{array}{ccc} 0 & 1 & 0\\
                                                                   0 & 0 & 1\\
                                                                   1 & 0 & 0\\
                                     \end{array} \right ) ,\mbox{ }
\Omega_{2} = \left ( \begin{array}{ccc} \exponential^{2 \pi i /3} & 0 & 0\\
                                        0 & \exponential^{-2 \pi i /3} & 0\\
                                                                  0 & 0 & 1\\
                                     \end{array} \right ) .
\eeq

Note that on the lattice the link variables, $U_{\mu}(\hat{n})$ ($\mu
= 1, 2, 3, 4$ is the Lorentz index and $\hat{n}$ is the position of a
lattice site),  obey the boundary condition
\beq
\label{eq:TBC_gauge}
U_\mu( \hat{n} + \hat \nu L_\nu/a )
= \Omega_\nu U_\mu( \hat{n} ) \Omega_\nu^\dagger\,.
\eeq

Fermions must also obey some specific boundary conditions in order to
maintain gauge invariance and single-valuedness of the action. In
particular it is well known that the number of fermion flavours have
to be an integer multiple of the rank of the gauge
group~\cite{Parisi:1984cy}. The different flavours (usually called
``smells'') transform one in the other under translations of a full
period of the torus according to
\beq
\psi^a_\alpha( \hat{n} + \hat{\nu} L_{\nu}/a) = e^{i\pi/3} \Omega_\nu^{ab}
\psi^b_\beta (\hat{n}) \left ( \Omega_\nu \right )^\dagger_{\beta
  \alpha} ,
\eeq
where $\nu=1,2$ and Latin ($a,b$) and Greek ($\alpha,\beta$) indices
are colour and smell indices, respectively. 
The factor $\exponential^{i\pi/3}$ is introduced to lift the
zero-momentum modes in these directions. 
In the directions $\nu = 3, 4$, we impose PBC, $\psi(
\hat{n}+ \hat{\nu} L_{\nu}/a ) = \psi( \hat{n} )$. 

Using TBC in lattice calculations leads to various advantages.
It removes the toron configurations.  Therefore in the
weak-coupling regime, the power laws in the coupling in finite-volume perturbation
theory are the same as those in the infinite-volume, continuum
case~\cite{GonzalezArroyo:1981vw, Coste:1985mn, GonzalezArroyo:1988dz}.
This
boundary condition also lifts the zero-momentum modes, making it
possible to perform simulations at vanishing fermion mass.

\subsection{The gradient flow and the renormalised coupling constant}
\label{sec:GF_coupling}
In recent years the Yang-Mills gradient
flow~\cite{Narayanan:2006rf,Luscher:2010iy} has become a standard
technique to define renormalised couplings (see~\cite{Ramos:2015dla}
for a recent review).

The basic idea is to add an extra coordinate to our gauge field, that
we will call ``flow time'' and denote it by $t$. The evolution of the
gauge field $B_\mu(t,x)$ with respect to the flow time is given by the
Yang-Mills gradient flow equation,
\begin{equation}
  \label{eq:flow}
  \frac{d B_\mu(t,x)}{dt} = D_{\mu}G_{\mu\nu}(t,x)\,,
\end{equation}
where $G_{\mu\nu}$ is the field strength associated with
$B_{\mu}(t,x)$, and $D_\mu = \partial_\mu + [B_\mu,\cdot]$ the
corresponding covariant derivative. The initial condition
for Eq.~(\ref{eq:flow}) is given by $B_\mu(t,x)|_{t=0} = A_\mu(x)$, where
$A_\mu(x)$ is the fundamental gauge field. It has been proven to all
orders in perturbation theory that gauge invariant composite
observables made of the flow field $B_\mu(t,x)$ are automatically
renormalised for $t>0$~\cite{Luscher:2011bx}. In
particular Ref.~\cite{Luscher:2010iy} suggests using the action density to
define a renormalised coupling at a scale $\mu = 1/\sqrt{8t}$,
\begin{equation}
  g^{2}(\mu) = \frac{16\pi^2}{3}t^2\langle E(t)\rangle\,,
\end{equation}
with
\begin{equation}
  E(t) = -\frac{1}{2}{\rm tr}(G_{\mu\nu}G_{\mu\nu})\,.
\label{eq:energy_density}
\end{equation}

In the context of finite size scaling, the renormalisation scale is
identified with the linear size of the box $\mu = \frac{1}{\sqrt{8t}}=
1/c_\tau L$ where $c_\tau = \sqrt{8t}/L$ is a constant that defines
our renormalisation 
scheme. Several finite volume renormalisation schemes have been
defined by using different boundary conditions:
periodic~\cite{Fodor:2012td}, Schr\"odinger functional~\cite{Fritzsch:2013je}, open-SF~\cite{Luscher:2014kea}, and more
directly related with this work, twisted boundary
conditions~\cite{Ramos:2014kla}.  

There is quite some freedom when translating the flow equation
Eq.~\eqref{eq:flow} to the lattice. Since the r.h.s. of
Eq.~\eqref{eq:flow} is just the gradient of the Yang-Mills action, a
straightforward option consists in evolving the lattice gauge links
$V_{\mu}(t,x)$ according to the gradient of a lattice
action\footnote{For the precise definition of the Lie-algebra valued
  derivative $\partial_{x,\mu}$ see~\cite{Luscher:2010iy}.
}
\beq
 \frac{\partial V_{\mu} (t,x)}{\partial t} = - g_{0}^{2} \left \{
   \partial_{x,\mu} S_{{\mathrm{latt}}} \left [ V_{\mu} \right
   ]\right \} V_{\mu} (t,x) , \mbox{ }\mbox{ } V_{\mu} (0,x) = U_{\mu}
 (x) .
\label{eq:WF_equation}
\eeq
Moreover one has also the freedom to choose amongst different
discretisations to define the observable $E(t)$, the most popular one
being the clover and the plaquette. In general when evaluating the
coupling these two choices of discretisations (i.e. the lattice flow
equation and the observable) are, together with the action chosen to
produce the configurations, the three sources of cutoff
effects~\cite{Fodor:2014cpa,Ramos:2014kka}. Although recently a
discretisation of the flow equation and observables free of $\mathcal
O(a^2)$ effects has been proposed~\cite{Ramos:2015baa}, in this work
we have used the Wilson flow (i.e. Eq.~\eqref{eq:WF_equation} with the
Wilson plaquette action for $S_{\rm latt}$), and two different
discretisations of the observable (clover and plaquette).

In full glory our coupling reads,
\beq
\label{eq:GF_scheme_def}
 \bar{g}^{2}_{\rm latt} (\beta,\hat{L}) = \hat{{\mathcal{N}}}^{-1} (c_{\tau}, a/L) t^{2}
 \la E^{({\rm latt})}(a,t)\ra \big{|}_{t = c_{\tau}^{2} L^{2} /8 } ,
\eeq
where
\beq
 \beta = \frac{6}{g_{0}^{2}}, \mbox{ } \hat{L} = L/a ,
\label{eq:beta_and_Lhat}
\eeq
$E^{({\rm latt})}(a,t)$ is the discretised version of $E(t)$ defined
in Eq.~(\ref{eq:energy_density}), and the constant $\hat{\mathcal N}(c_\tau, a/L)$ has been computed
to tree-level with our choice of discretisations
(see~\cite{Ramos:2014kla} for the concrete expressions).

\subsection{The step-scaling method}
\label{sec:SS_method}
In identifying gauge theories that are viable for the walking
technicolour scenario, one may have to follow the 
evolution of the running coupling constant for a large range of scales.  
Although the lattice regularisation introduces both IR (the volume)
and ultraviolet (the lattice spacing) cut-off scales which normally
differ by only a factor of twenty to one hundred,
such a task is made possible by employing the step-scaling method.  In the following, we briefly review this method, 
in the context of our calculation of the coupling constant in the GF scheme.

%
%
  
To apply the step-scaling technique in our work, we first compute the coupling constant, $\bar{g}_{\rm latt}$, 
defined in Eq.~(\ref{eq:GF_scheme_def}).  This technique relies on the
use of the lattice size as the renormalisation scale.  To ensure that
this is feasible, one has to make certain that in addition to any
intrinsic scale in the theory under investigation, $L$ is the only
length scale that is being introduced in the analysis.  This means
that we have to work with fixed $c_{\tau}$\footnote{Different choices
of $c_{\tau}$ correspond to various renormalised schemes defined using
the Yang-Mills gradient flow.}, and remove the effects of
the lattice spacing.

By performing this computation at various values of $\hat{L}$ 
and $\beta$ [Eq.~(\ref{eq:beta_and_Lhat})], we can tune these input parameters to determine the renormalised coupling 
that does not depend on the lattice spacing,
\beq
\label{eq:continuum_g_input}
 g_{\rm GF} \left ( L \right ) = \bar{g}_{\rm cont} \left ( L \right ) = \bar{g}_{\rm latt} \left ( \beta_{1}, L/a_{1} \right ) = \bar{g}_{\rm latt} \left ( \beta_{2}, L/a_{2} \right ) = 
  \ldots = \bar{g}_{\rm latt} \left ( \beta_{n_{0}}, L/a_{n_{0}} \right ) .
\eeq
In practice, one simulates at fixed values of $\hat{L}$.  Since the
lattice spacing, $a$, is determined by the bare coupling, $\beta$ (or
$g_{0}^{2}$), the above equation describes a procedure of adjusting
$a$ to achieve a constant physical length scale, $L$ at various chosen
$\hat{L}$ in the simulations.
This tuned coupling, $g_{\rm GF} \left ( L \right ) = \bar{g}_{\rm cont}
\left ( L \right )$, is the ``input coupling'' in the determination of
the step-scaling function.  
Following the conventional notation, we
define the ``input variable'', $u$, as
\beq
 u \equiv \bar{g}_{{\rm cont}}^{2}\left ( L \right ) =
 g^{2}_{{\rm GF}}\left ( L \right ) .
\label{eq:def_u}
\eeq
In practice, the tuning in Eq.~(\ref{eq:continuum_g_input}) is carried out by choosing a few (as denoted by $n_{0}$) values of $\hat{L}$, and adjusting $\beta$ accordingly to achieve the above condition.  After this tuning, $\bar{g}_{\rm cont}(L)$ does not depend the lattice spacing, therefore must be renormalised at the length scale
$L$.  In this work, our simulations for this ``tuning procedure'' have been performed at 
\beq
\label{eq:choices_of_L_over_a}
 L/a = 8, 10 ,12.
\eeq
The details of the tuning for $\beta$ will be given in Sec.~\ref{sec:beta_interpolation}.

%
%

Using the $n_{0}$ values of $(\beta, L/a)$, as tuned {\it via} Eq.~(\ref{eq:continuum_g_input}), 
we then calculate the step-scaling function,
\beq
\label{eq:lattice_step_scaling_function}
 \Sigma \left ( \beta_{i}, L/a_{i}, u, s \right ) \equiv \bar{g}^{2}_{\rm latt}\left . 
   \left ( \beta_{i}, sL/a_{i}\right ) \right |_{u = \bar{g}^{2}_{\rm latt} \left ( \beta_{i}, L/a_{i}\right )} , \mbox{ } i = 1, 2,\ldots , n_{0} ,
\eeq
where $s$ is the step size.   With these $n_{0}$ results for $\Sigma$ at the same physical
volume, $L$, but different lattice spacings, we can perform the continuum extrapolation for the step-scaling function,
\beq
\label{eq:step_scaling_function}
 \sigma \left ( u, s \right ) \equiv \bar{g}^{2}_{\rm cont} \left
   . \left ( sL \right ) \right |_{u=\bar{g}^{2}_{\rm cont}\left ( L \right )} 
  = \lim_{a \rightarrow 0} \Sigma \left ( \beta_{i}, L/a_{i}, u, s \right ) .
\eeq
We set the step size $s=2$ in this work. Given the choices of $\hat{L}$ in Eq.~(\ref{eq:choices_of_L_over_a}), we then have to compute the step-scaling function at
\beq
\label{eq:sL_over_a}
 sL/a = 16, 20, 24 .
\eeq
With these three values of $\hat{L}$, two different discretisations
for the energy density in Eq.~(\ref{eq:energy_density}) are implemented.
This allows us to investigate the
systematic effects arising from the continuum extrapolation.  We 
will give details of this issue in Sec.~\ref{sec:continuum_extrap}.

%
%

To make the presentation clear, we define
\beq
\label{eq:sigma_at_s2}
 \sigma \left ( u \right ) \equiv \sigma \left ( u, s = 2 \right ) .
\eeq
The step-scaling function is simply the renormalised coupling, therefore its value depends on the choice of renormalisation scheme.  A more suitable approach in demonstrating the existence of the IRFP is through the calculation of the ratio, 
\beq
\label{eq:r_sigma_def}
 r_{\sigma} \left ( u \right ) \equiv \frac{\sigma \left ( u \right
 )}{u} = \frac{g^{2}_{\rm GF}(2L)}{g^{2}_{\rm GF}(L)}.
\eeq
This ratio becomes one when the $\beta$-function vanishes.  The existence of zeros of the $\beta{-}$function is scheme-independent, although the values of the coupling at these zeros are not.
In order to confirm that an asymptotically-free gauge theory contains an IRFP, we have to 
show that in this theory $r_{\sigma}(u)$ is one at both
ultraviolet (UV) and IR regimes, while being positive in between.
This demonstrates that when increasing the length scale, the theory
flows out of the vicinity of the UV Gaussian fixed point, and then flows
towards the IRFP at strong coupling.

\subsection{Simulation parameters and the raw data}
\label{sec:simulation_parameters}
In this work, we perform simulations using the Wilson
plaquette gauge action, and unimproved massless staggered fermions.  As
discussed
in Sec.~\ref{sec:TBC}, the number of flavours in our
calculation has to be a multiple of $N_{c} = 3$, as a result of the
introduction of the smell degrees of freedom.  Staggered fermions
contain four tastes, making it suitable for the lattice computation of
SU(3) gauge theory with twelve flavours.  

%
%

Our simulations are performed with the standard HMC algorithm.  The molecular-dynamics evolution is carried out 
using the Omelyan integrator with multi-time
steps~\cite{Sexton:1992nu,Hasenbusch:2001ne}.  For the inversion of
the fermion matrix, we use the biCGstab solver with the tolerance set
to be $10^{-16}$.  The numerical accuracy for the Metropolis tests is $10^{-24}$.
A large portion of our
computation, including all the $\hat{L} \ge 20$ simulations, were performed using Graphics Processing Units (GPU's).

%
%

To implement the step-scaling method for computing the running
coupling constant, as discussed in Sec.~\ref{sec:SS_method}, we
carry out simulations at the lattice volumes,
\beq
\label{eq:lattice_volumes}
 L/a = \hat{L} = 8, 10, 12, 16, 20, 24.
\eeq
For each volume, we perform lattice calculations at several values
of the bare coupling constant.  As mentioned in the
Introduction,  the slow-running behaviour of the coupling in
SU(3) gauge theory with twelve flavours results in the demand for
controlling the statistical error to the subpercentage level.  Furthermore, this
behaviour makes it necessary to perform computations at a large range of
bare coupling (lattice spacing), in order to trace the renormalised 
coupling from the UV to
the IR regimes.  We have the input bare
coupling, $g_{0}^{2}$, mostly in the range,
\beq
  4.1 \le \beta = \frac{6}{g_{0}^{2}} \le 20.0 ,
\label{eq:beta_values}
\eeq
in the simulation.  For each $\hat{L}$, we perform computations at 
20 to 34 choices of $\beta$, as summarised in
Table~\ref{tab:bare_g_summary}.   In our analysis procedure, we have
to interpolate the renormalised couplings extracted from simulations 
at different bare couplings, as detailed in
Sec.~\ref{sec:beta_interpolation}.  Having a large number of data
points for this interpolation is essential for obtaining statistically
independent results in the step-scaling study, as discussed at the end
of Sec.~\ref{sec:results_and_discussion}.
\begin{table}[t]
\begin{center}
\begin{tabular}{|c|c|c|c|c|c|c|}
\hline\hline
$\hat{L}$ & 8 & 10 & 12 & 16 & 20 & 24\\
\hline
\# of $\beta$ values & 34 & 32 & 34 & 31 & 21 & 20\\
\hline
Minimal $g_{0}^{2}$ & 0.298 & 0.298 & 0.298 & 0.298 & 0.120 & 0.300\\
\hline
Maximal $g_{0}^{2}$ & 1.460 & 1.456 & 1.463 & 1.449 & 1.446 &
1.442\\
\hline\hline
$(\bar{g}^{2}_{{\rm latt}})_{{\rm max}}^{{\rm clover}, 0.325}$ & 6.502(12) & 6.555(12) & 6.747(11) & 6.544(9) & 6.557(9) & 6.532(9)\\
\hline
$(\bar{g}^{2}_{{\rm latt}})_{{\rm max}}^{{\rm clover},0.350}$ & 6.510(14) & 6.553(14) & 6.748(13) & 6.541(10) & 6.560(11) & 6.527(11)\\
\hline
$(\bar{g}^{2}_{{\rm latt}})_{{\rm max}}^{{\rm clover},0.375}$ & 6.489(16) & 6.530(15) & 6.734(14) & 6.524(13) & 6.549(12) & 6.507(13)\\
\hline
$(\bar{g}^{2}_{{\rm latt}})_{{\rm max}}^{{\rm clover},0.400}$ & 6.438(18) & 6.482(17) & 6.697(16) & 6.486(15) & 6.518(14) & 6.465(15)\\
\hline
$(\bar{g}^{2}_{{\rm latt}})_{{\rm max}}^{{\rm clover},0.450}$ & 6.243(21) &  6.296(21) & 6.534(21) & 6.327(18) & 6.375(18) & 6.296(20)\\
\hline
$(\bar{g}^{2}_{{\rm latt}})_{{\rm max}}^{{\rm clover},0.500}$ & 5.926(25) & 5.984(24) & 6.241(26) & 6.043(22) & 6.105(24) & 5.998(25)\\
\hline\hline
$(\bar{g}^{2}_{{\rm latt}})_{{\rm max}}^{{\rm plaq},0.325}$ & 7.024(12) & 6.813(12) & 6.911(11) & 6.623(9) & 6.606(9) & 6.565(9)\\
\hline
$(\bar{g}^{2}_{{\rm latt}})_{{\rm max}}^{{\rm plaq},0.350}$ & 6.920(14) & 6.761(13) & 6.884(13) & 6.608(11) & 6.601(11) & 6.555(10)\\
\hline
$(\bar{g}^{2}_{{\rm latt}})_{{\rm max}}^{{\rm plaq},0.375}$ & 6.820(16) & 6.705(15) & 6.849(14) & 6.581(12) & 6.584(12) & 6.531(12)\\
\hline
$(\bar{g}^{2}_{{\rm latt}})_{{\rm max}}^{{\rm plaq},0.400}$ & 6.715(18) & 6.633(17) & 6.797(16) & 6.536(14) & 6.550(14) & 6.487(14)\\
\hline
$(\bar{g}^{2}_{{\rm latt}})_{{\rm max}}^{{\rm plaq},0.450}$ & 6.459(21) & 6.421(21) &6.618(21)  & 6.370(18) & 6.402(18) & 6.315(20)\\
\hline
$(\bar{g}^{2}_{{\rm latt}})_{{\rm max}}^{{\rm plaq},0.500}$ & 6.112(25) & 6.096(24) & 6.318(26) & 6.083(22) & 6.130(23) & 6.016(25)\\
\hline\hline
\end{tabular} 
\caption{Summary of choices of $\beta$ ($g_{0}^{2}$) values in the
simulation.  Also shown are the corresponding largest
$\bar{g}^{2}_{{\rm latt}}$ at each $\hat{L}$,
from both the clover and the plaquette discretisations at
some representative values of $c_{\tau}$, with the notation $(\bar{g}^{2}_{{\rm latt}})_{{\rm max}}^{{\rm discretisation},c_{\tau}}$.}
\label{tab:bare_g_summary}
\end{center}
\end{table}
In this table, we also list the corresponding results of $\bar{g}^{2}_{{\rm latt}}$ at the largest $g_{0}^{2}$
for a few selections of $c_{\tau}$.  The values of $\bar{g}^{2}_{{\rm latt}}$
and $g_{0}^{2}$ differ significantly in the strong-coupling regime.
On the other hand, as a consequence of asymptotic freedom, 
at $g_{0}^{2} \sim 0.3$, we observe that the bare
and the renormalised coupling strengths are similar.    

To avoid strong cut-off effects, one should work with large enough $c_{\tau}$,
such that $\bar{g}^{2}_{{\rm latt}}$ decreases with increasing $c_{\tau}$ at
fixed $g_{0}^{2}$ and $\hat{L}$.  For this purpose,
it is clear from Table~\ref{tab:bare_g_summary}  that our analysis has
to be carried out 
with data obtained at $c_{\tau} \ge 0.375$.   We implement the
Yang-Mills gradient flow for many values of $c_{\tau}$.
However, in order to illustrate our study in this work, it suffices to
present results at
\beq
c_{\tau} = 0.375, 0.400, 0.450, 0.500 ,
\label{eq:ctau_for_this_paper}
\eeq
in the remaining of this paper.

%
%

As discussed in Refs.~\cite{Lin:2012iw, Aoyama:2011ry},  in lattice
calculations of SU(3) gauge theory with twelve massless flavours, the
Markov chains can exhibit tunnelling behaviour amongst local minima of
the effective potential.   Such behaviour may lead to artificially long
autocorrelation time, and should be monitored carefully.   To reduce
the probability of having this tunnelling in our simulations, we thermalise the Markov chains by starting
from a configuration with zero fermion mass, and
\bea
\label{eq:conf_ntv}
U_{3}( \hat{n}_{1},\hat{n}_{2},\hat{n}_{3}=1,\hat{n}_{4}) &=& e^{ - 2 i \pi / 3 } 
~,~U_{4}(\hat{n}_{1},\hat{n}_{2},\hat{n}_{3},\hat{n}_{4}=1) = e^{ + 2 i \pi / 3 }~,\\ \nonumber
U_{\mu}(\hat{n}_{1},\hat{n}_{2},\hat{n}_{3},\hat{n}_{4}) &=& 1 \rm{~~elsewhere } .
\eea
This forces the system to be at the true
vacuum, in which Polyakov loops in the untwisted directions contain
non-vanishing imaginary parts~\cite{Aoyama:2011ry}.    The above
prescription also produces the largest gap in the vicinity of zero
modes in the fermion matrix.   We observe that the autocorrelation time
for the renormalised coupling, $\bar{g}^{2}_{{\rm latt}}$,  remains at
about 20 to 100 HMC trajectories, and it
increases with $c_{\tau}$.  
We have been able to obtain data with
statistical uncertainties for $\bar{g}^{2}_{{\rm latt}}$ below
$0.5\%$, even for the largest
flow time that corresponds to $c_{\tau} = 0.5$ in this work.
This is achieved with about 100,000
HMC trajectories.

%
%

It is also important that we implement our simulations away from
artificial bulk phases in the lattice theory~\cite{Cheng:2011ic}.  For
this purpose, we have checked the plaquette expectations values and
confirmed that all our computations were performed in the
weak-coupling phase.

In Appendix~\ref{sec:raw_data}, we tabulate all our raw data for the renormalised couplings, extracted
using both clover and plaquette discretisations, at $c_{\tau} = 0.5$.
Lattice data at other values of $c_{\tau}$ can be obtained from the
authors upon request.

\section{Analysis details for the step-scaling study}
\label{sec:analysis}
In this section, we give the details of our analysis
procedure.   As shown in Table~\ref{tab:bare_g_summary} 
in the last section, we perform simulations at 172 combinations of
the bare coupling, $g_{0}$, and the lattice size, $\hat{L}$.  It can
also be seen in this table that typical statistical errors for our raw
data at the largest couplings are between $\sim 0.2\%$ ($c_{\tau} =
0.375$) to $\sim 0.45\%$ ($c_{\tau} = 0.500$).  This feature is common in
all values of $(g_{0}, \hat{L})$ in this work.   Such error budget is
achieved by having at least $\sim 100,000$ HMC trajectories, 
followed by carrying out measurement every 50 to 200 trajectories,
and creating 100 to 200 bins for data at each simulation.   One
thousand bootstrap samples are created for our analysis.

\subsection{Interpolation in the bare coupling}
\label{sec:beta_interpolation}
In principle, the step-scaling method described in
Sec.~\ref{sec:SS_method} has to be implemented by tuning
the bare coupling, on the $L/a = 8, 10, 12$ lattices,  
to achieve the condition in Eq.~(\ref{eq:continuum_g_input}).
However, in this work we aim at tracing the coupling constant over a
large range of scale, while computing the ratio, $r_{\sigma}(u)$,
at many values of the input renormalised coupling, $u$, in the
process.  This renders the time-consuming tuning procedure impractical.  
In view of this, we make use of an interpolation method as the
substitution for the tuning of the bare coupling.  Namely,  for each $L/a$ we perform
simulations at many values of $\beta$ in a wide range, and then
obtain $\bar{g}_{{\rm latt}} (\beta, L/a)$ at other bare couplings in
this range through interpolation.  In the rest of this
section, we discuss this procedure in detail.

\begin{figure}[t]
\begin{center}
\scalebox{0.45}{\includegraphics{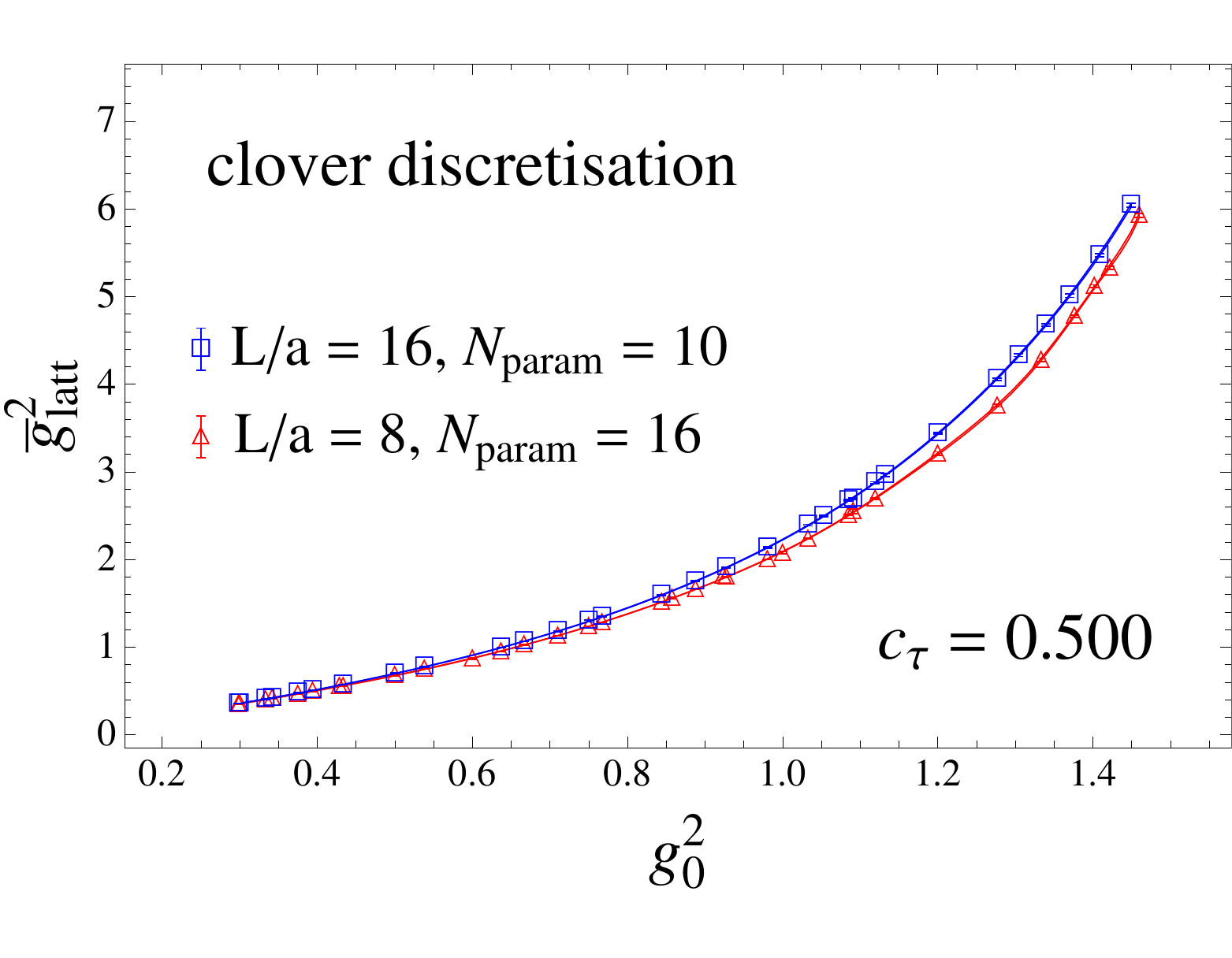}}
\hspace{0.5cm}
\scalebox{0.45}{\includegraphics{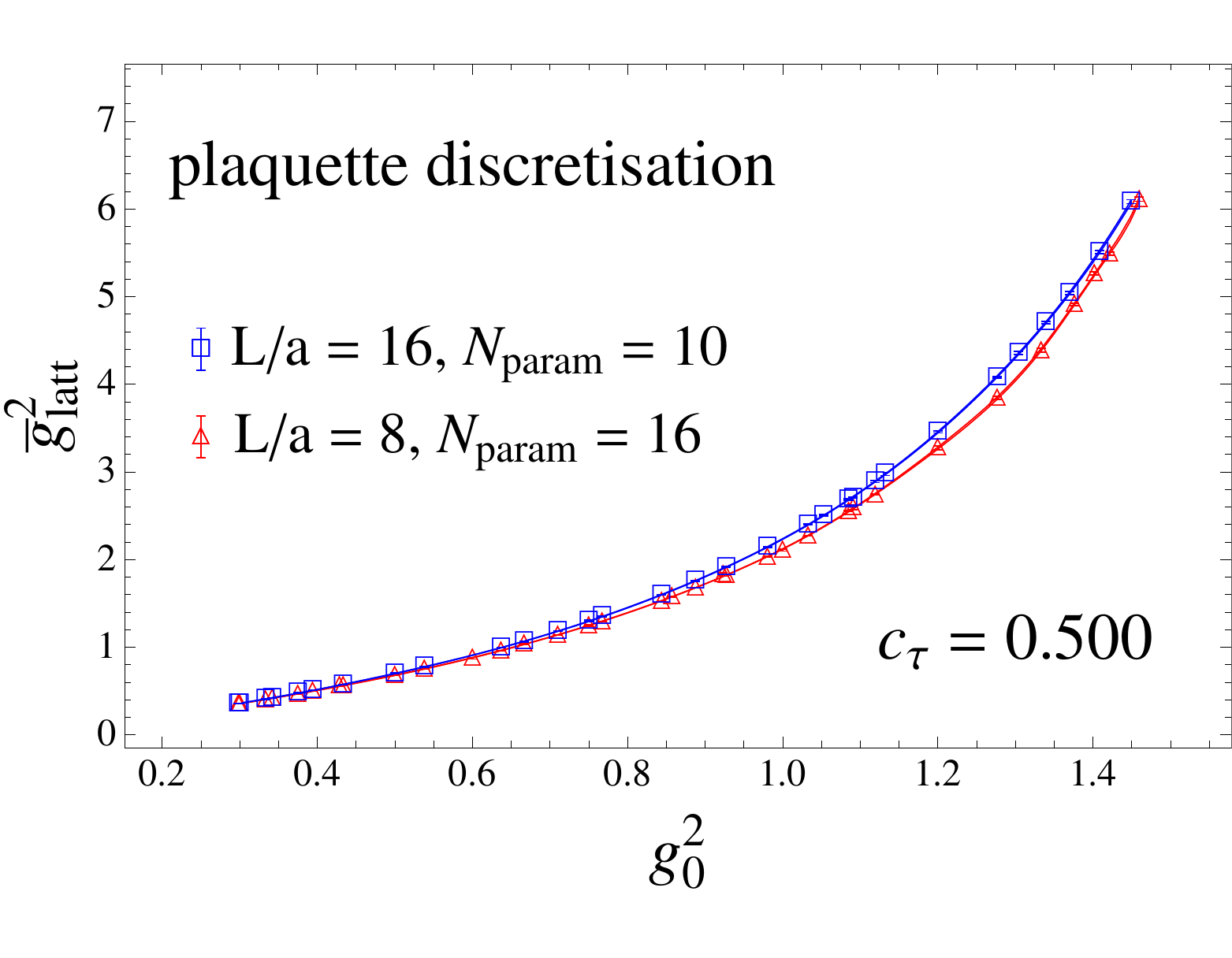}}\\
\scalebox{0.45}{\includegraphics{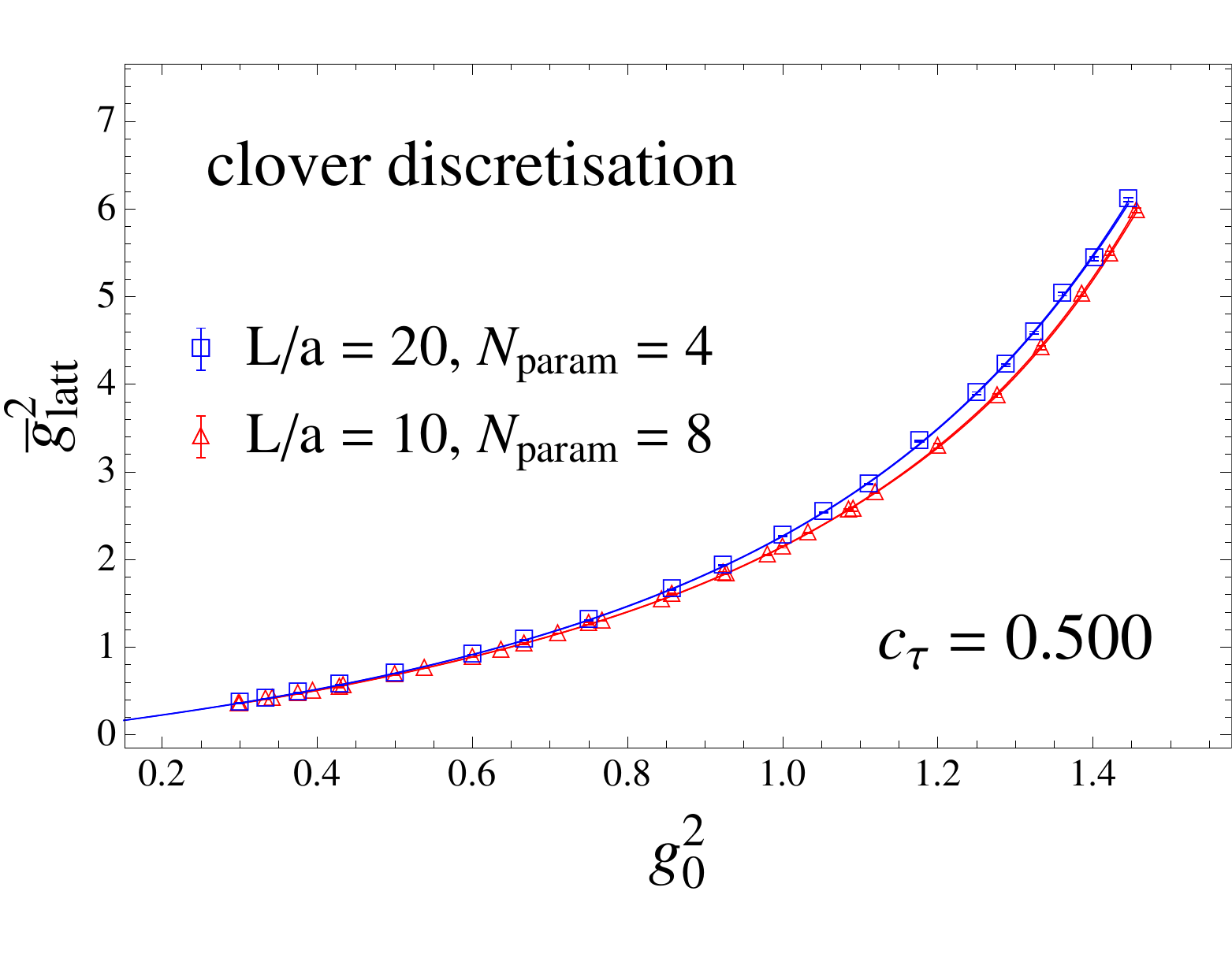}}
\hspace{0.5cm}
\scalebox{0.45}{\includegraphics{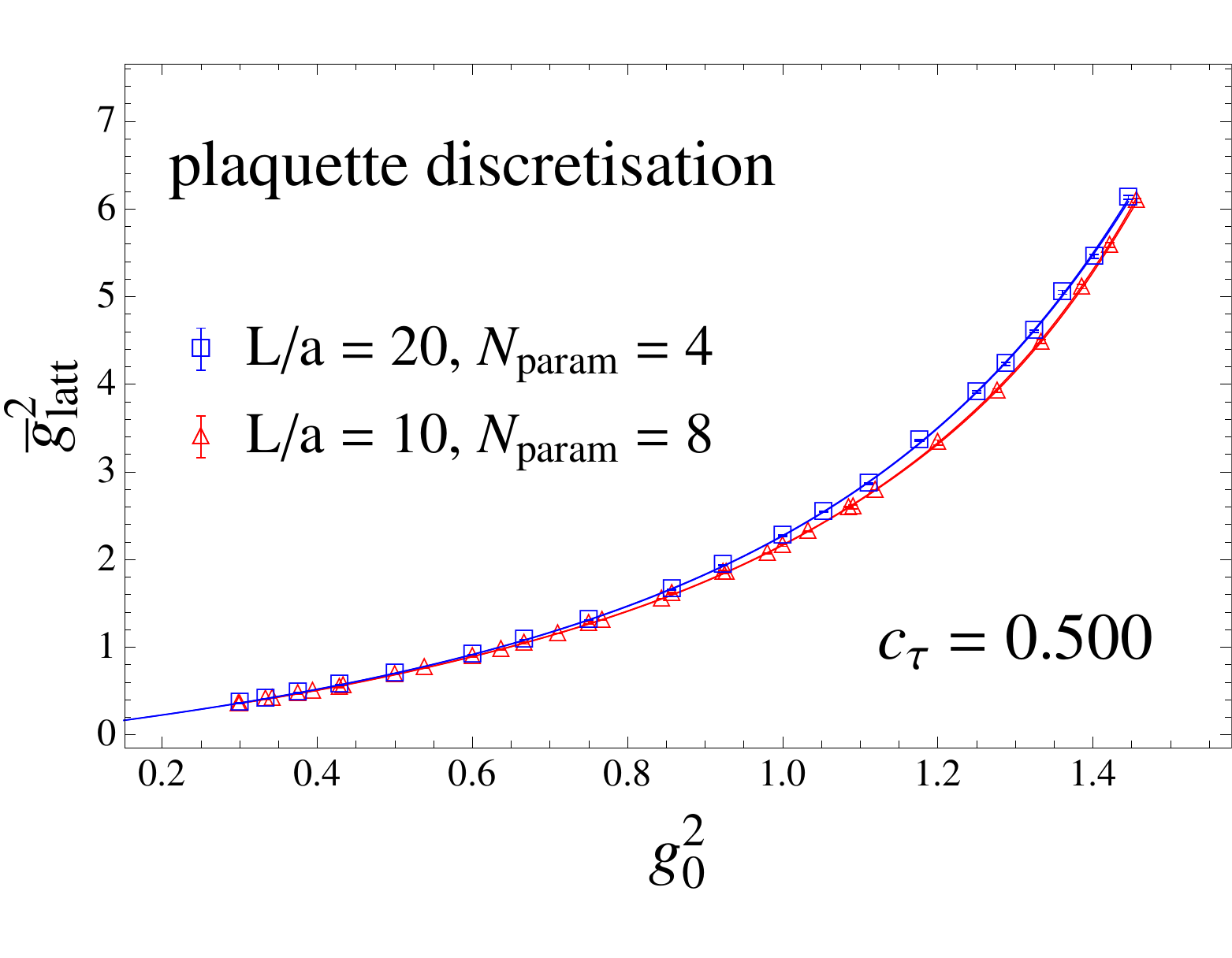}}\\
\scalebox{0.45}{\includegraphics{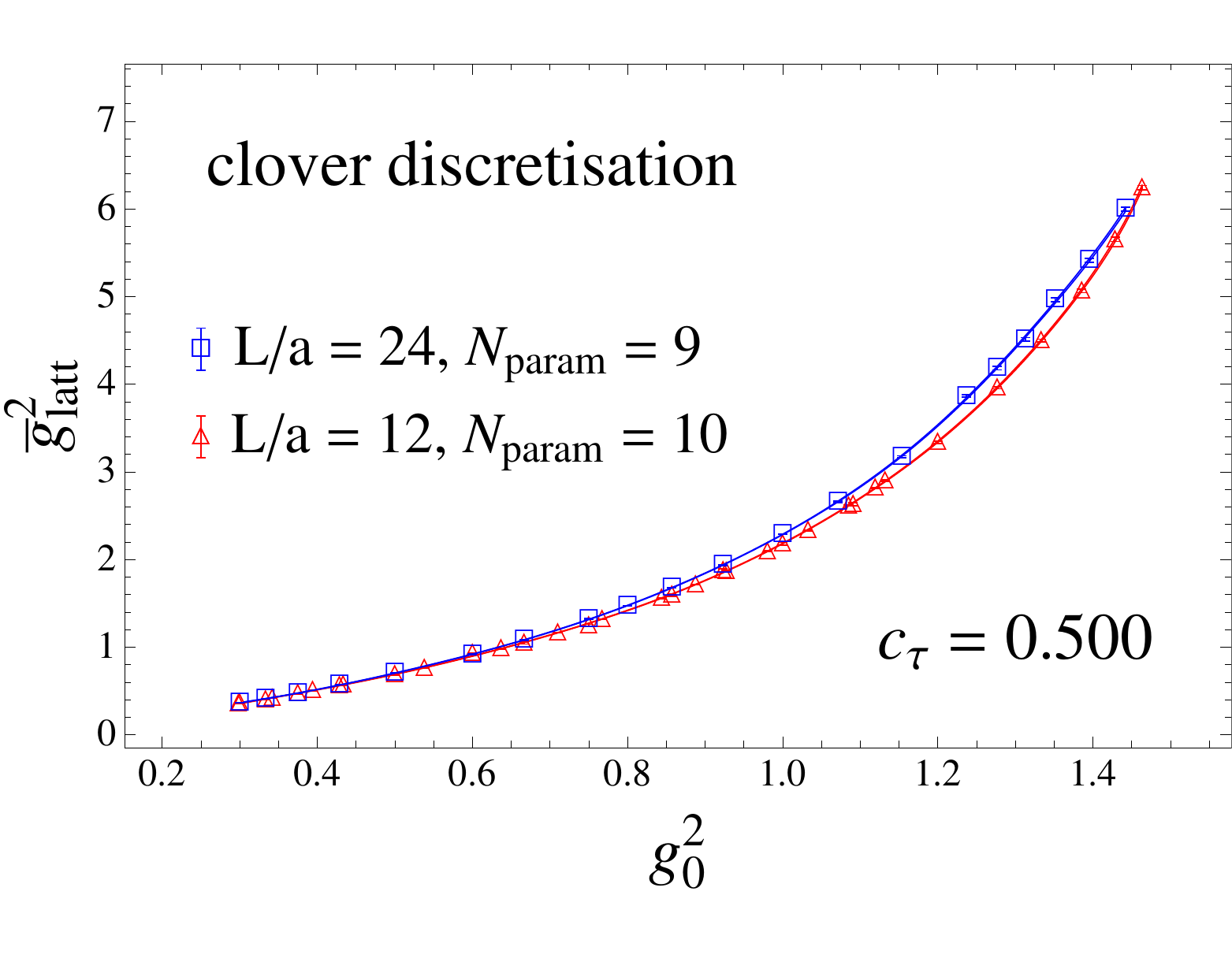}}
\hspace{0.5cm}
\scalebox{0.45}{\includegraphics{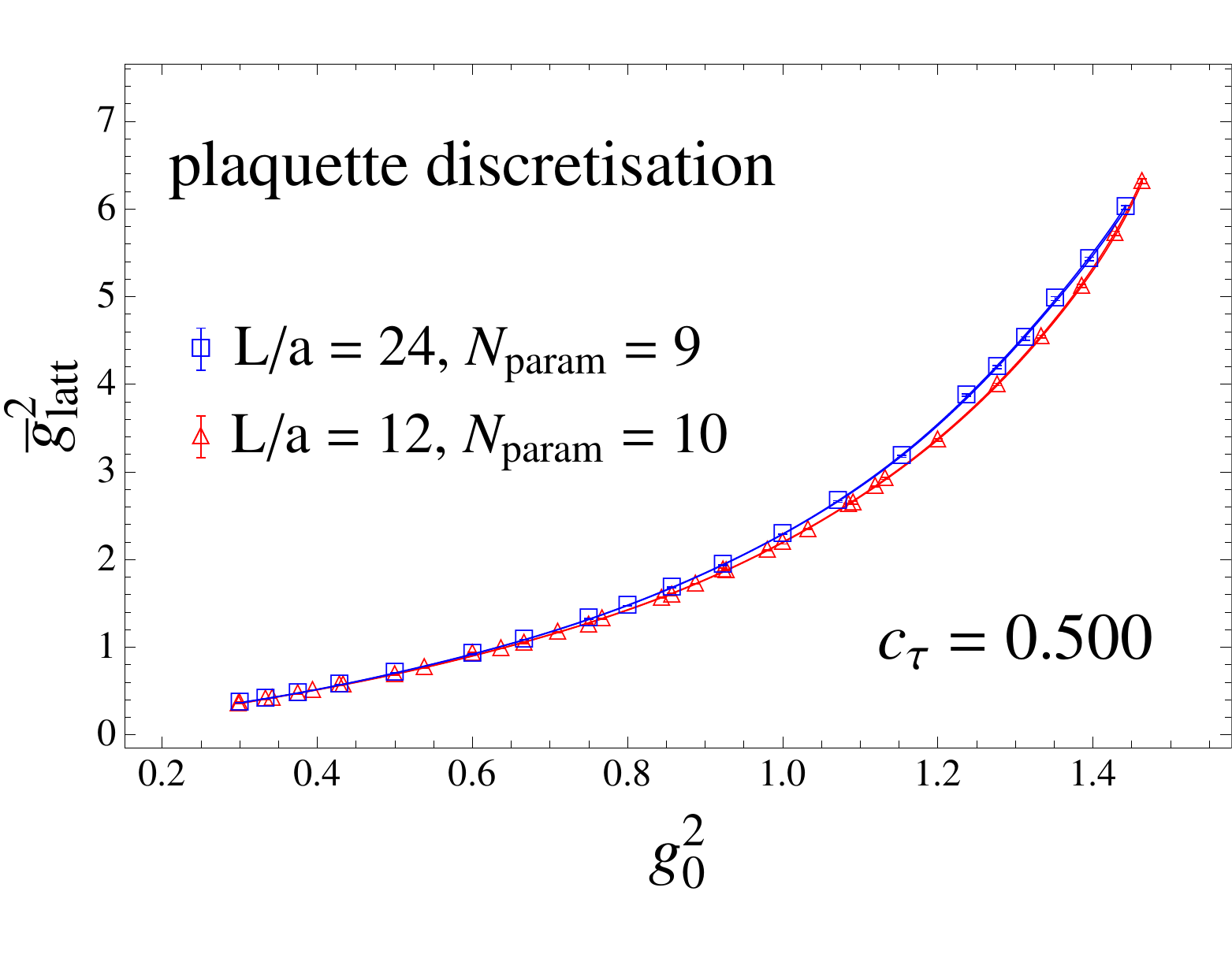}}\\
\caption{
The renormalised coupling, $u_{\rm latt} = \bar{g}^{2}_{{\rm latt}}$,
from the simulations on the volumes $L/a=8,10,12,16,20,24$ at
$c_{\tau} = 0.5$.  The raw data are displayed by points with error bars. 
Fit functions, Eq.~(\ref{eq:non_decr_poly}),  are shown as curves.
Notice that the curves contain statistical errors in the fits. 
}
\label{fig:coupling}
\end{center}
\end{figure}

The feature of the input $\beta{-}$values in our simulations is given
in Table~\ref{tab:bare_g_summary} of
Sec.~\ref{sec:simulation_parameters}.  
Since the chosen range of bare coupling constant straddles between the 
perturbative and non-perturbative domains,  it
is challenging to perform the interpolation in $\beta$ using a
well-inspired function.   We begin by noting that in the perturbative
(large${-}\beta$) regime,  one-loop approximation
has to be applicable.  Therefore, at fixed $L/a$,
\beq
 u_{\rm latt} (\beta, L/a) \equiv \bar{g}^{2}_{\rm latt} (\beta, L/a) \approx
 \frac{6}{\beta} = g^{2}_{0}, {\rm ~at~high~} \beta, 
\label{eq:high_beta_interp}
\eeq
with $g_{0}$ being the bare gauge coupling.
Equation~(\ref{eq:high_beta_interp}) also leads to the motivation for
using polynomials in $1/\beta$ to
carry out the bare-coupling interpolation.  In this work, we perform
simulations for many values of $\beta$
at each $L/a$.  This makes it possible to use high-degree polynomials
which can in principle result in good fits.
Nevertheless, having such high-degree polynomials in the interpolation
procedure can introduce artificial oscillatory behaviour of the fit
function, known as the Runge phenomenon.   In order to avoid this
artefact, we first note that the renormalised coupling, $u_{\rm latt}$,
should be non-decreasing in $1/\beta$ at fixed $L/a$ in this work,
since our simulations are all performed in the weak-coupling phase of
the lattice theory.
In view of this, we can impose the
non-decreasing constraint on the bare-coupling interpolation and use
the function,
\beq
u_{\rm latt} = f(u_{0}) = \int_{0}^{u_{0}} du \mbox{ }\left (
  \sum_{m=0}^{N_{\rm deg}} c_{m} u^{m} \right )^{2} = \sum_{n=0}^{N_{h}}
h_{n} u_{0}^{n},
\label{eq:non_decr_poly}
\eeq
where 
\beq
u_{0} \equiv \frac{1}{\beta} =
  \frac{g^{2}_{0}}{6} .
\eeq
It can be seen in 
Fig.~\ref{fig:coupling}, where we plot $u_{{\rm latt}}$
against $g^{2}_{0}$, that this constraint is well-justified.
Combining 
Eqs.~(\ref{eq:high_beta_interp}) and (\ref{eq:non_decr_poly}), we
further impose
\beq
 h_{0} = 0 , \mbox{ } h_{1}=6 \mbox{ } ({\rm then~}c_{0} = \sqrt{6}) .
\label{eq:non_decr_constraint}
\eeq
%
%
This condition results in the number of fit parameters,
\beq
 N_{\rm param} = N_{\rm deg} = \frac{N_{h}-1}{2} ,
\label{eq:nparam_ndeg}
\eeq
with $N_{\rm deg}$ and $N_{h}$ defined in
Eq.~(\ref{eq:non_decr_poly}).

In addition to reducing the artificial oscillatory behaviour in the
bare-coupling interpolation, the non-decreasing polynomial fit
function has the advantage that its inverse is singled-valued.
This single-valuedness is an essential requirement in the
implementation of the step-scaling method.  It is also a necessary
consequence resulting from the fact that all our simulations are
performed in the weak-coupling phase of the lattice theory.

Figure~\ref{fig:coupling} shows the bare-coupling
interpolation at $c_{\tau} = 0.5$ using
Eq.~(\ref{eq:non_decr_poly}).
In this figure, $N_{\rm param}$ is fixed to result in the best $\chi^{2}/{\rm
  d.o.f.}$ volume by volume (see Table~\ref{tab:chisq_beta_interpolation}).  The interpolation
is performed with uncorrelated fits.  It is clear from this plot that
the resulting curves are smooth and they explain the data well.  We
find the same behaviour for all our other choices of $c_{\tau}$.
In Table~\ref{tab:chisq_beta_interpolation}, we give the values of
$N_{\rm param}$ corresponding to the best
$\chi^{2}/{\rm d.o.f.}$ in these fits.   In this interpolation
procedure, we observe that $\chi^{2}/{\rm d.o.f.}$ is close to unity for all the
optimal fits.
\begin{table}[t]
\begin{center}
\begin{tabular}{|c||c|c|c|c|}\hline
$L/a$ & \multicolumn{4}{|c|}{optimal $N_{\rm param}$ for clover discretisation}\\ \hline
    & $c_{\tau} = 0.375$ & $c_{\tau} = 0.400$ & $c_{\tau} = 0.450$ & $c_{\tau} = 0.500$\\ \hline
8  & 16 & 16 & 16 & 16\\ \hline
10 & 7 &  7 & 8 & 8\\ \hline
12 & 10 &  10 & 10 & 10\\ \hline
16 & 10 &  10 & 10 & 10\\ \hline
20 & 4 &  4 & 4 & 4\\ \hline
24 & 10 &  9 & 9 & 9\\ \hline
\end{tabular}
\hspace{10mm}
\begin{tabular}{|c||c|c|c|c|}\hline
$L/a$ & \multicolumn{4}{|c|}{optimal $N_{\rm param}$ for plaquette discretisation}\\ \hline
    & $c_{\tau} = 0.375$ & $c_{\tau} = 0.400$ & $c_{\tau} = 0.450$ & $c_{\tau} = 0.500$\\ \hline
8  & 16 & 16 & 16 & 16\\ \hline
10 & 7 &  7 & 8 & 8\\ \hline
12 & 10 &  10 & 10 & 10\\ \hline
16 & 10 &  10 & 10 & 10\\ \hline
20 & 4 &  4 & 4 & 4\\ \hline
24 & 10 &  9 & 9 & 9\\ \hline
\end{tabular}
\caption{Optimal choices of $N_{\rm param}$ ($N_{\rm param}=N_{\rm deg} = \frac{N_{h}
    -1}{2}$)  of the bare-coupling
  interpolation using Eq.~(\ref{eq:non_decr_poly}) for the clover
  (left) and the plaquette (right) discretisations.}
\label{tab:chisq_beta_interpolation}
\end{center}
\end{table}

The results of this $\beta{-}$interpolation will be used in the
step-scaling study of the renormalised coupling.  It is obvious that
this can lead to correlation amongst results of the step-scaling
function and $r_{\sigma}$ determined at different input couplings.
This issue will be addressed in Sec.~\ref{sec:results_and_discussion}.

\subsection{Continuum extrapolation}
\label{sec:continuum_extrap}
Through the bare-coupling interpolation procedure, we can achieve the
tuning of the input renormalised coupling in
Eq.~(\ref{eq:continuum_g_input}).  This allows us to compute the
corresponding lattice step-scaling function, $\Sigma(\beta, L/a, u, s=2)$, in
Eq.~(\ref{eq:lattice_step_scaling_function}).  The next step in the
analysis is the extrapolation of these lattice step-scaling functions
to the continuum limit, as indicated in
Eq.~(\ref{eq:step_scaling_function}).  Since our simulations are
performed using unimproved staggered fermions and the standard Wilson
plaquette gauge action, the lattice artefacts are polynomials\footnote{There can be logarithmic corrections which are
  normally difficult to determine numerically.} of
$(a/L)^{2}$.   In this work, we carry out computations at $\hat{L} =
8, 10, 12, 16, 20, 24$.  This allows us to determine $\Sigma(\beta, L/a, u, s=2)$ at
\beq
 \hat{L} = L/a \left ( 8, 10, 12 \right ) \longrightarrow
  2 \hat{L} = \left ( 16, 20, 24 \right ) ,
\eeq
then extrapolate to the continuum limit with the linear function
\beq
\label{eq:linear_continuum_extrap}
 \Sigma (\beta, L/a, u, s=2) = \sigma (u) + A_{l} (u) \left ( \frac{a}{L}
 \right )^{2} .
\eeq
Notice that this fit function is valid only when the effects of the
lattice spacing are governed by the Gaussian fixed point in
the UV.   On the other hand, when the theory is engineered to be close enough to an
IRFP, its scaling behaviour regarding the change of both the
lattice spacing and the finite volume must be completely determined by
the IR scale invariance.
In this situation, the ``Symanzik-type'' polynomial
extrapolation of
Eq.~(\ref{eq:linear_continuum_extrap}) is not applicable, and
alternative methods have to be adopted.   We will
discuss this issue in Sec.~\ref{sec:FSS}.

It is well known that the continuum extrapolation is the main source
of systematic errors in the step-scaling investigation of the coupling
constant.  To check that this procedure is implemented reliably in our work,
we make use of two discretisations, namely the clover and the
plaquette, to compute the energy density defined in
Eq.~(\ref{eq:energy_density}).   These two discretisations
contain different lattice artefacts.  On the
other hand, any result obtained with these methods should extrapolate to
the same continuum limit, if the discretisation effects are under
control.

\begin{figure}[t]
\includegraphics[scale=0.45]{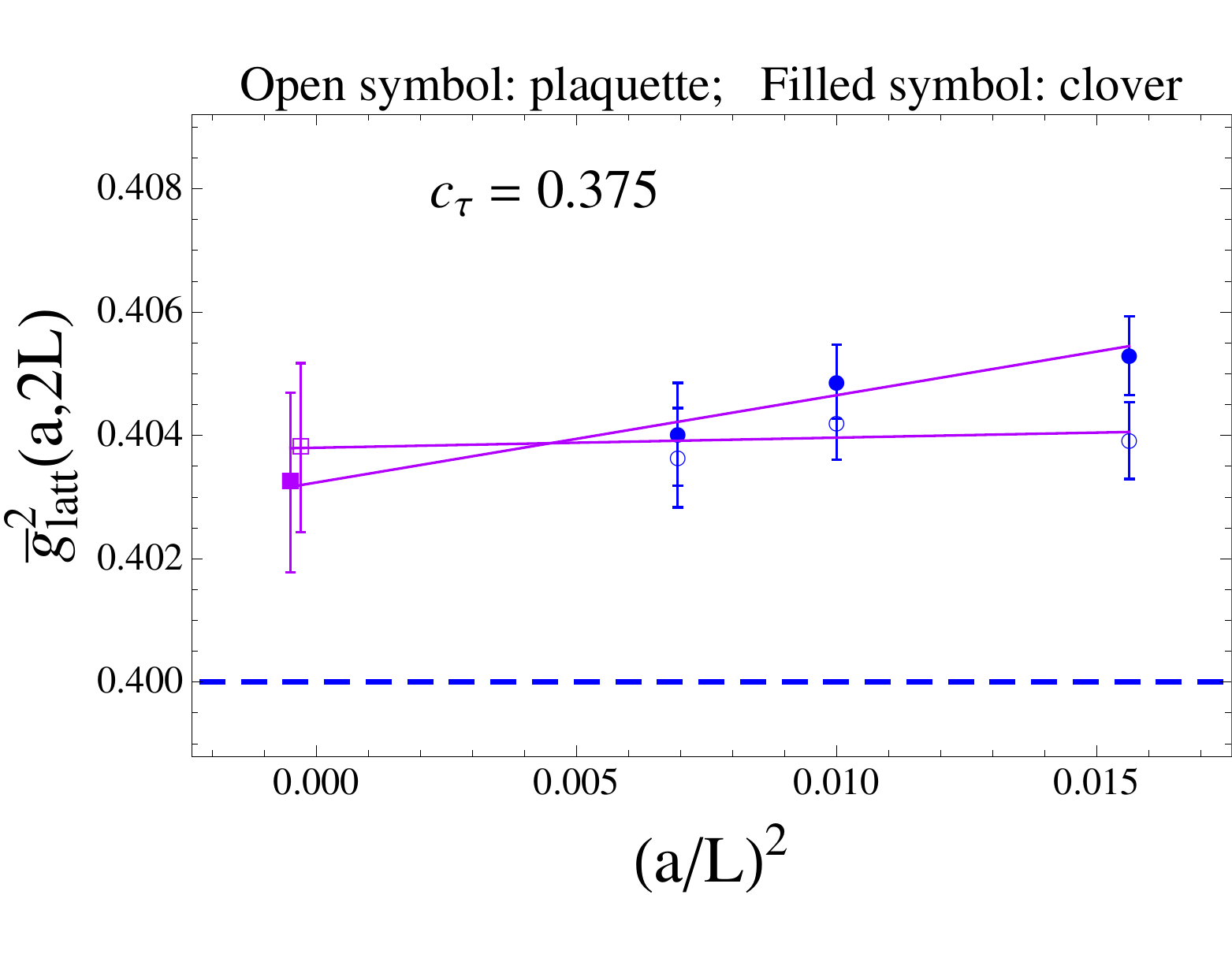}
\hspace{0.5cm}
\includegraphics[scale=0.45]{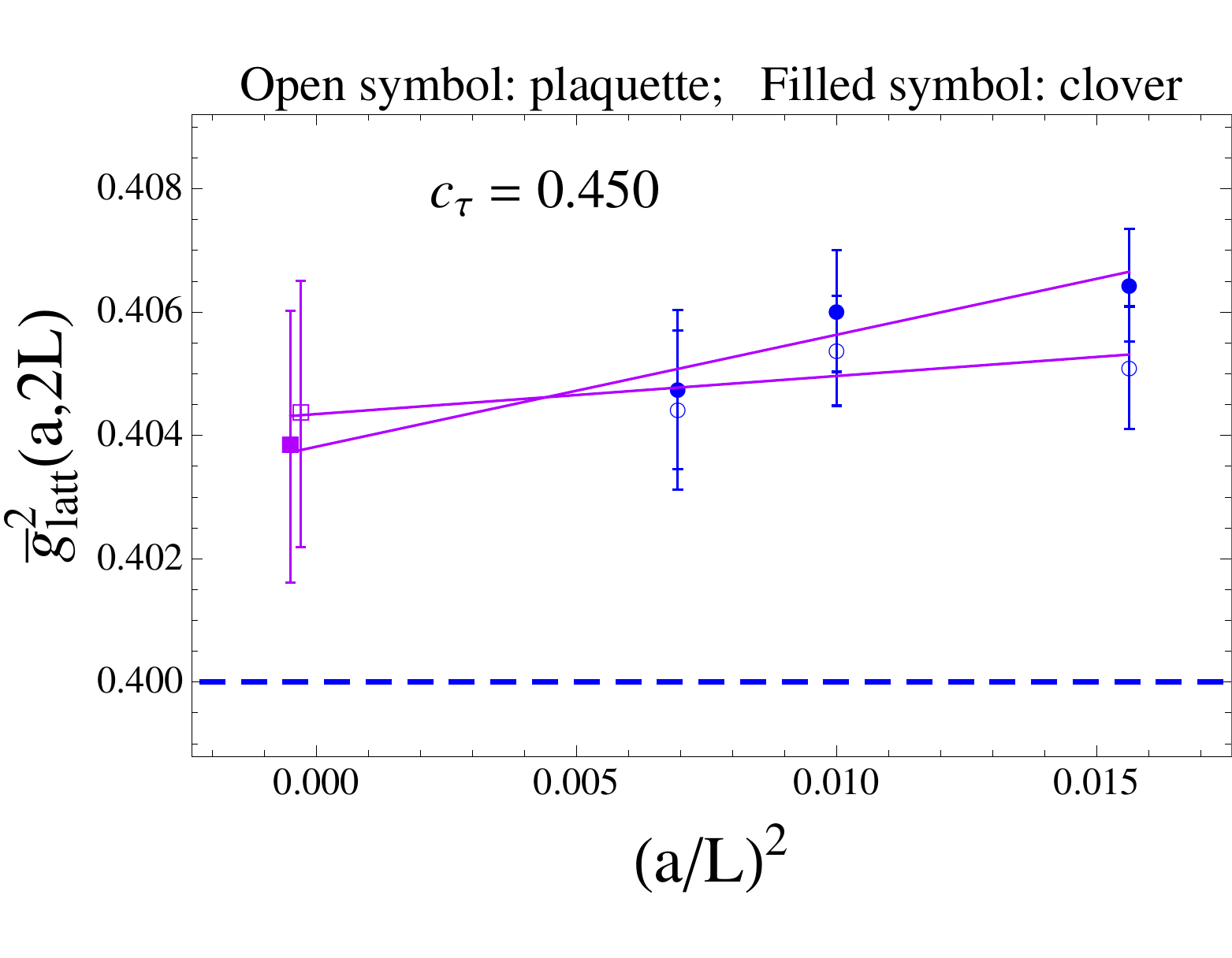}\\
\includegraphics[scale=0.45]{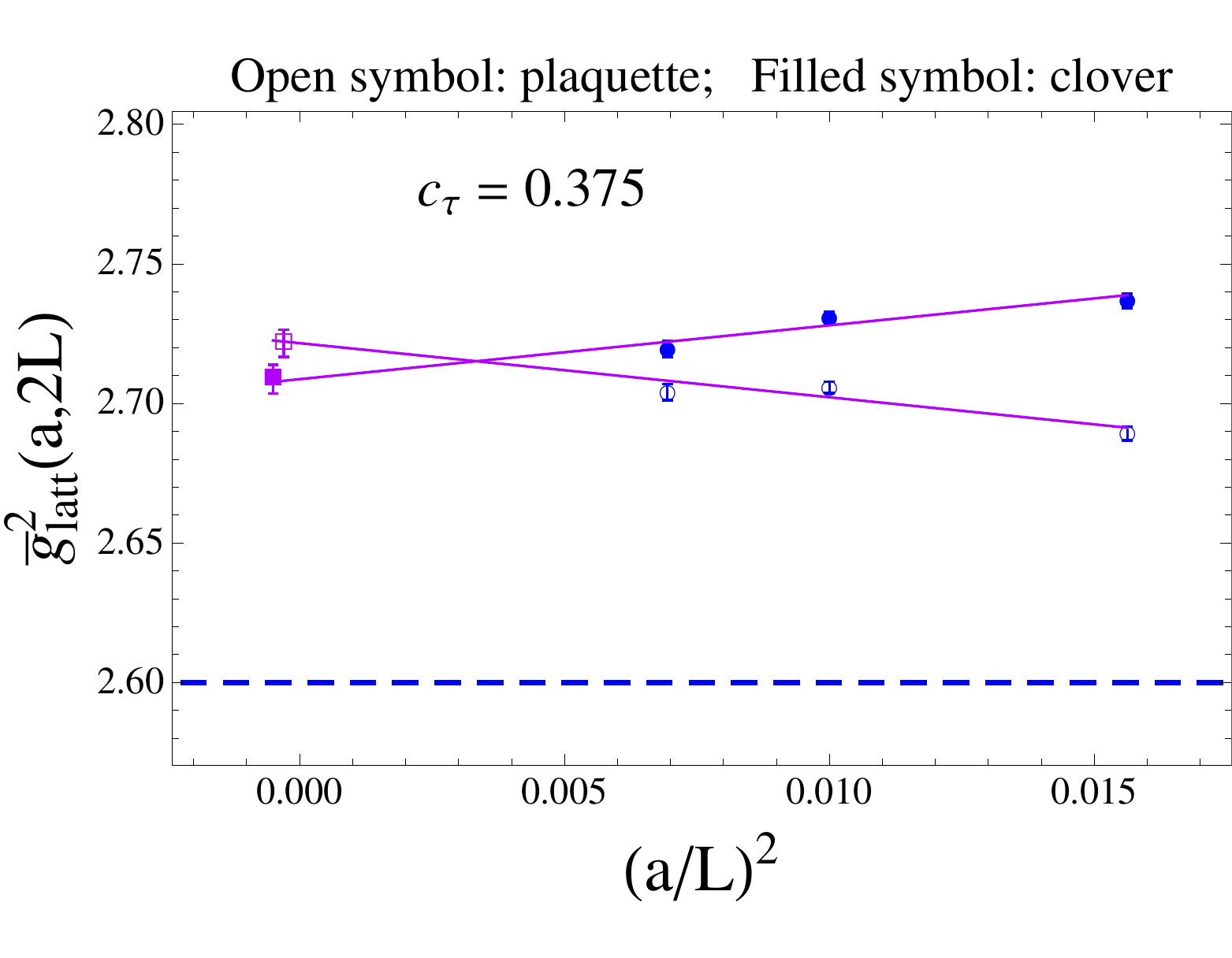}
\hspace{0.5cm}
\includegraphics[scale=0.45]{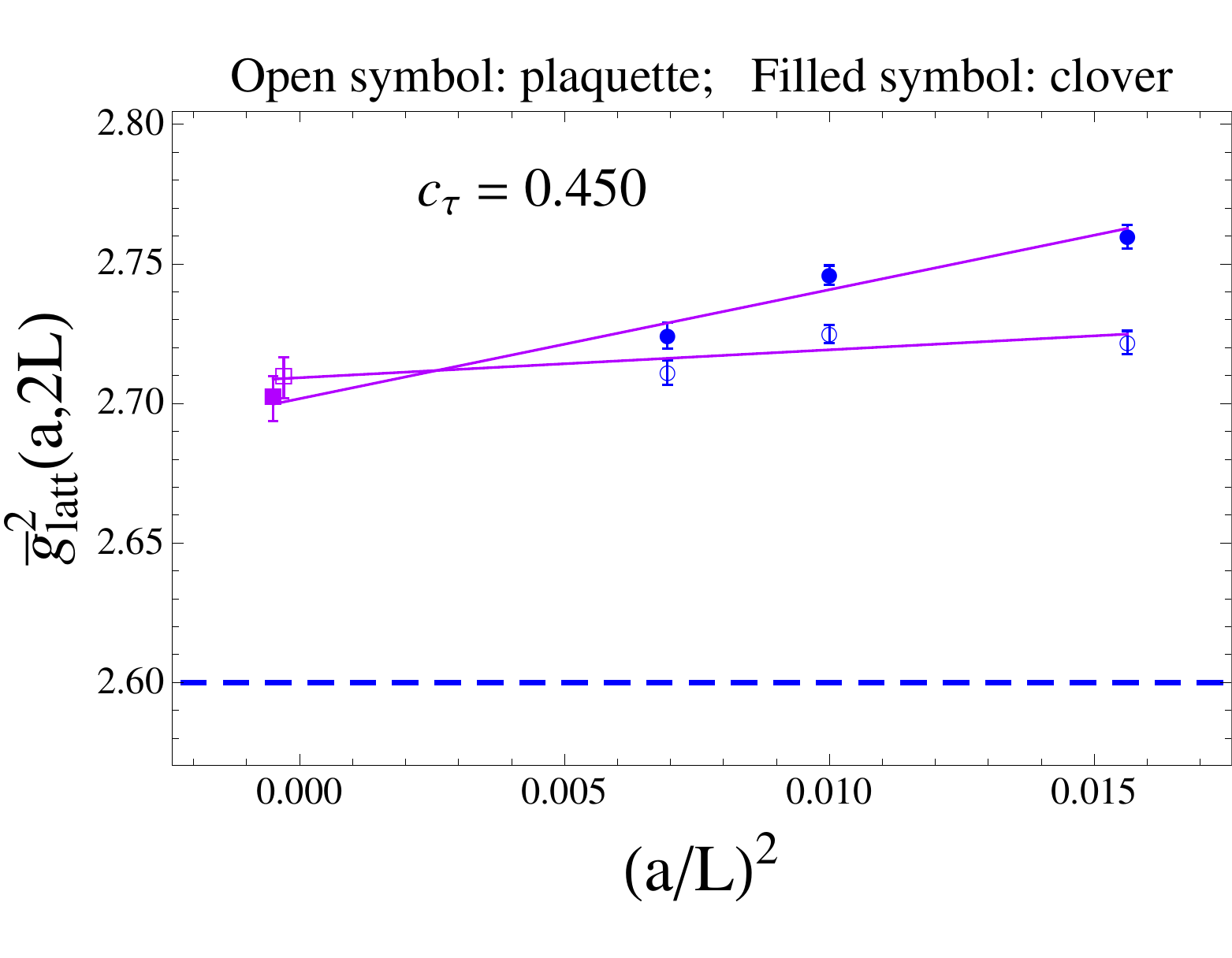}\\
\includegraphics[scale=0.45]{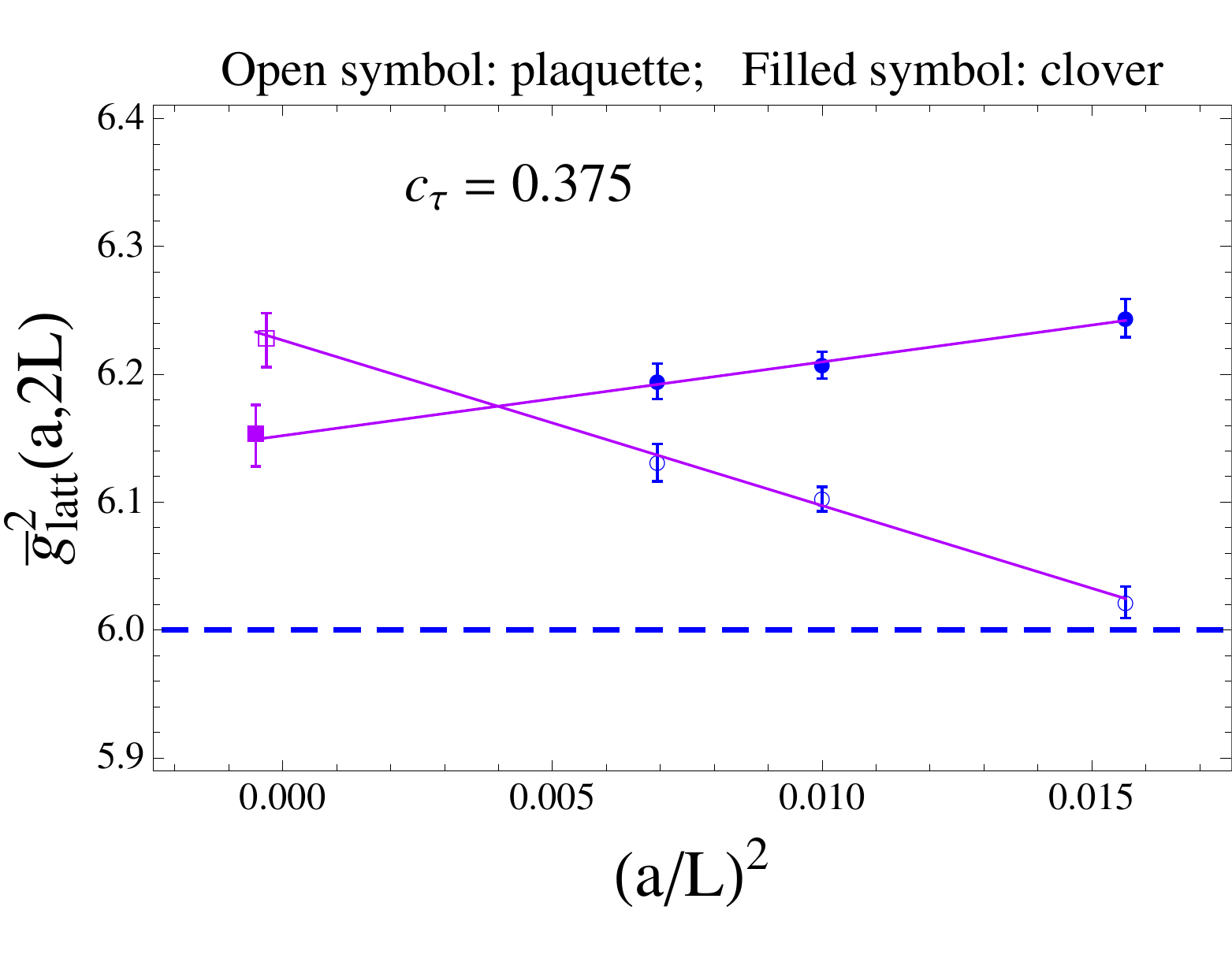}
\hspace{0.5cm}
\includegraphics[scale=0.45]{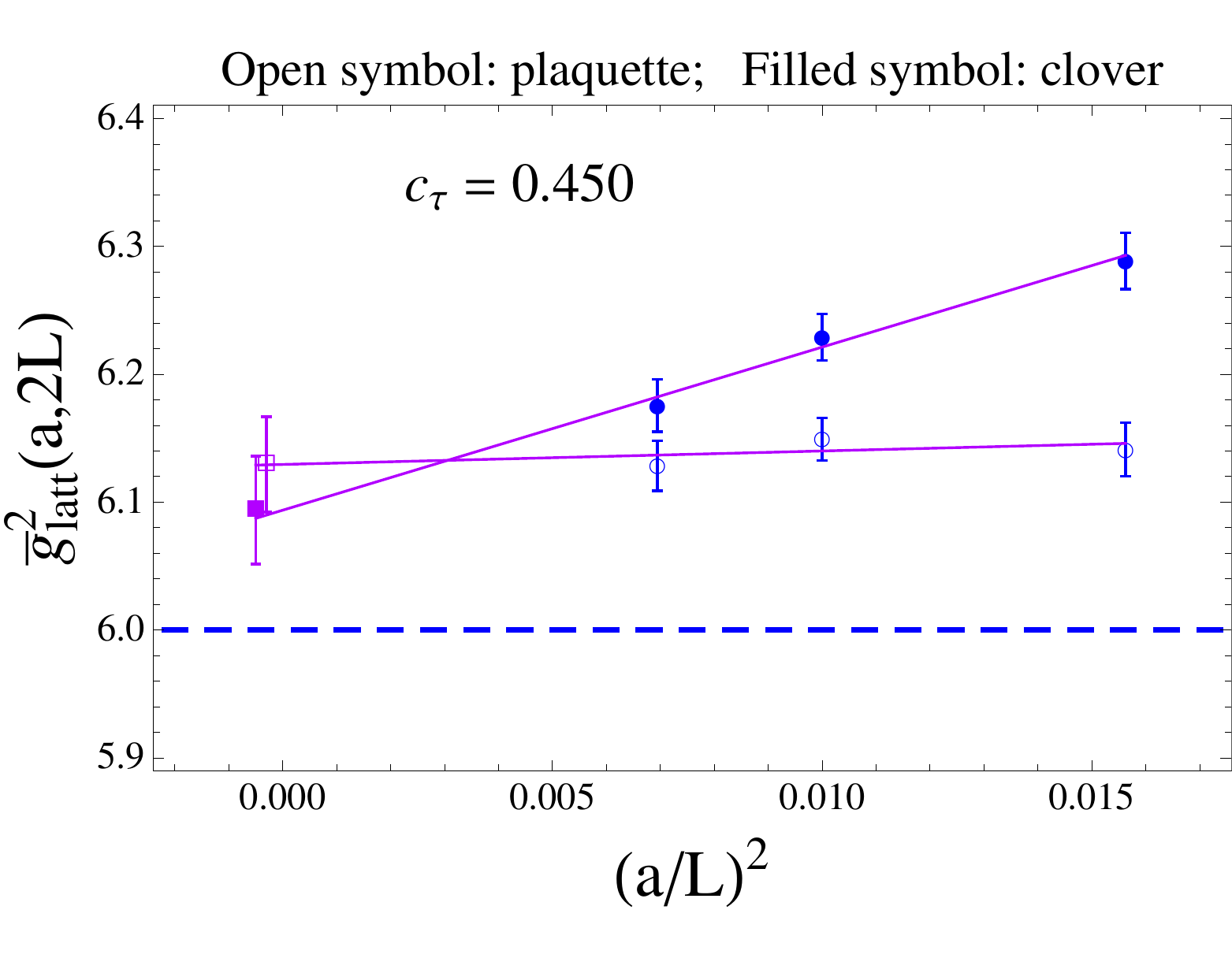}
\caption{Representative cases of the continuum extrapolation for the step-scaling functions
  with the procedure discussed in the main text.  
The dashed horizontal lines indicate the input
reference $\bar{g}^{2}_{{\rm cont}}(L)$, as tuned using 
Eq.~(\ref{eq:continuum_g_input}).  The data points at $a/L \not= 0$
are the lattice step-scaling functions defined in
Eq.~(\ref{eq:lattice_step_scaling_function}).}
\label{fig:ssf_continuum_extrap}
\end{figure}

Figure~\ref{fig:ssf_continuum_extrap} shows
representative plots of the continuum extrapolation at $c_{\tau} = 0.375$ and
$0.5$.   
We first notice that all the extrapolations are mild.   Even at
strong coupling, the change of $\bar{g}^{2}_{{\rm latt}} (g_{0}^{2},
2\hat{L})$ from our coarsest lattice ($\hat{L} = 8$) to the continuum
limit is at the level of a few percent.  This is partly because we
extract the renormalised coupling using a result from lattice perturbation
theory for the factor $\hat{{\mathcal{N}}}$ in Eq.~(\ref{eq:GF_scheme_def}).  
For the case of $c_{\tau} = 0.375$ at strong renormalised coupling,
the two discretisations do not lead to compatible results in the
continuum limit.
This renders the analysis unreliable in the IR regime.  We stress that all
fits for the continuum extrapolation produce good or acceptable $\chi^{2}/{{\rm
    d.o.f.}}$ (typically between 0 and 2) at this value of $c_{\tau}$, and $\bar{g}^{2}_{{\rm latt}} (g_{0}^{2},
2\hat{L})$ only weakly depends on $(a/L)^{2}$.   Nevertheless, this does
not mean that the procedure is under control.   We further comment
that this observation is made possible because our data are obtained
at small enough statistical errors.  
In the same figure, it is demonstrated that effects of the lattice artefacts are reduced
when $c_{\tau}$ is increased.   This is expected, since the smearing
radius of the gauge field grows with $c_{\tau}$.
For the case of $c_{\tau} = 0.5$, the
clover and the plaquette discretisations produce consistent results in
the continuum limit at all values of the coupling investigated in this project.
In the current work, this
extrapolation is under control, {\it i.e.}, the continuum-limit
results obtained from the two discretisation methods are compatible,
up to $g^{2}_{{\rm GF}} \sim 6$ when $c_{\tau} \ge 0.45$.  This can
also be seen clearly in Fig.~\ref{fig:ssf_continuum_extrap}.

\subsection{Results and discussion}
\label{sec:results_and_discussion}
We present results from our main analysis in this section.   As discussed
in Sec.~\ref{sec:beta_interpolation}, we perform the bare-coupling
interpolation using a non-decreasing polynomial function,
Eq.~(\ref{eq:non_decr_poly}), with the perturbation-theory constraint
in Eq.~(\ref{eq:non_decr_constraint}).   As already pointed out in
previous similar studies, this interpolation procedure is not
introducing significant systematic effects.   In this work we 
observe that this is also true in our analysis, by varying the numbers of
parameters reported in Table~\ref{tab:chisq_beta_interpolation}.
On the other hand,  we find that the systematic errors associate with the
continuum extrapolation can be significant.  Therefore we concentrate on the
discussion of this aspect of the error estimation in this section.

\begin{figure}[t]
\includegraphics[scale=0.45]{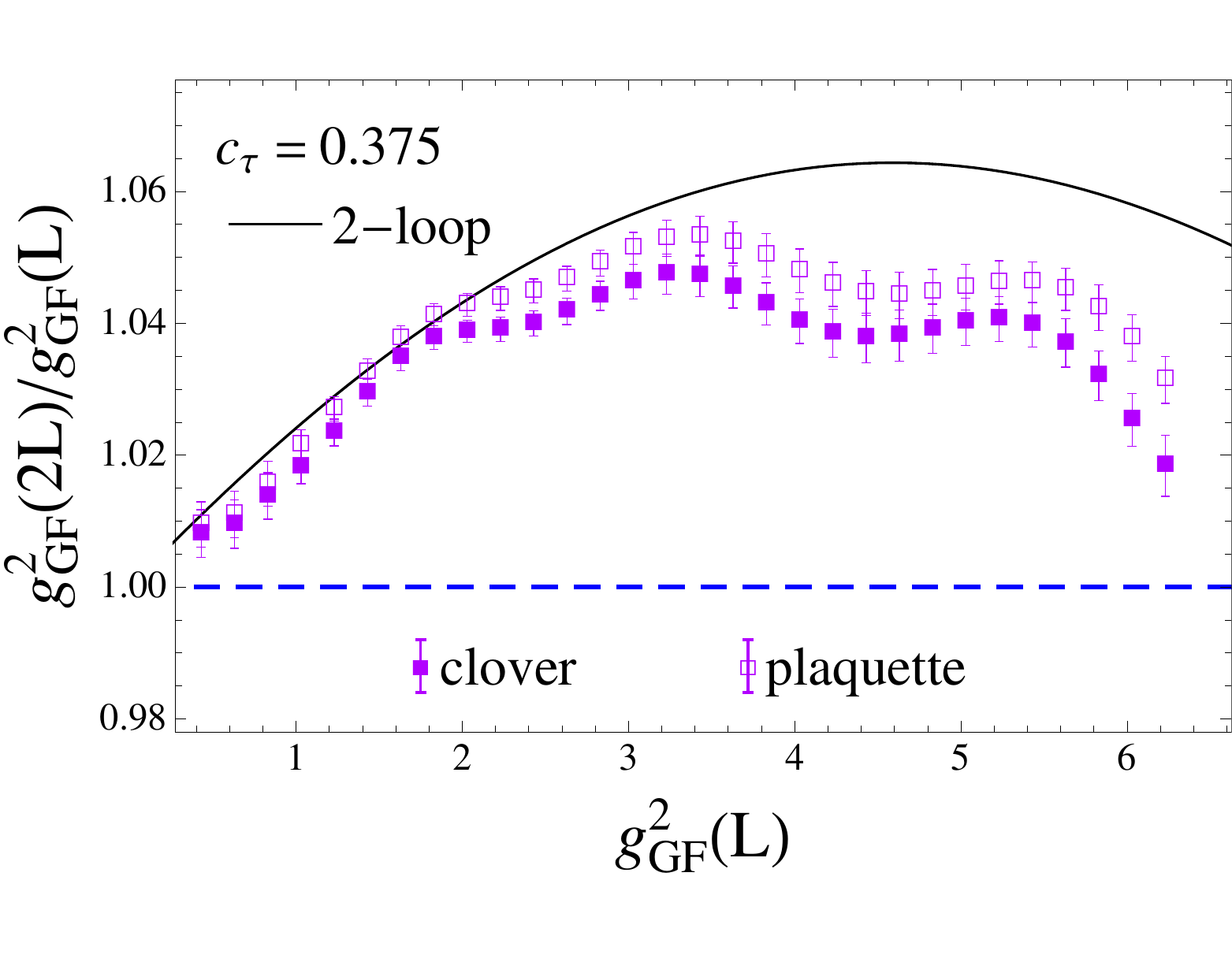}
\hspace{0.5cm}
\includegraphics[scale=0.45]{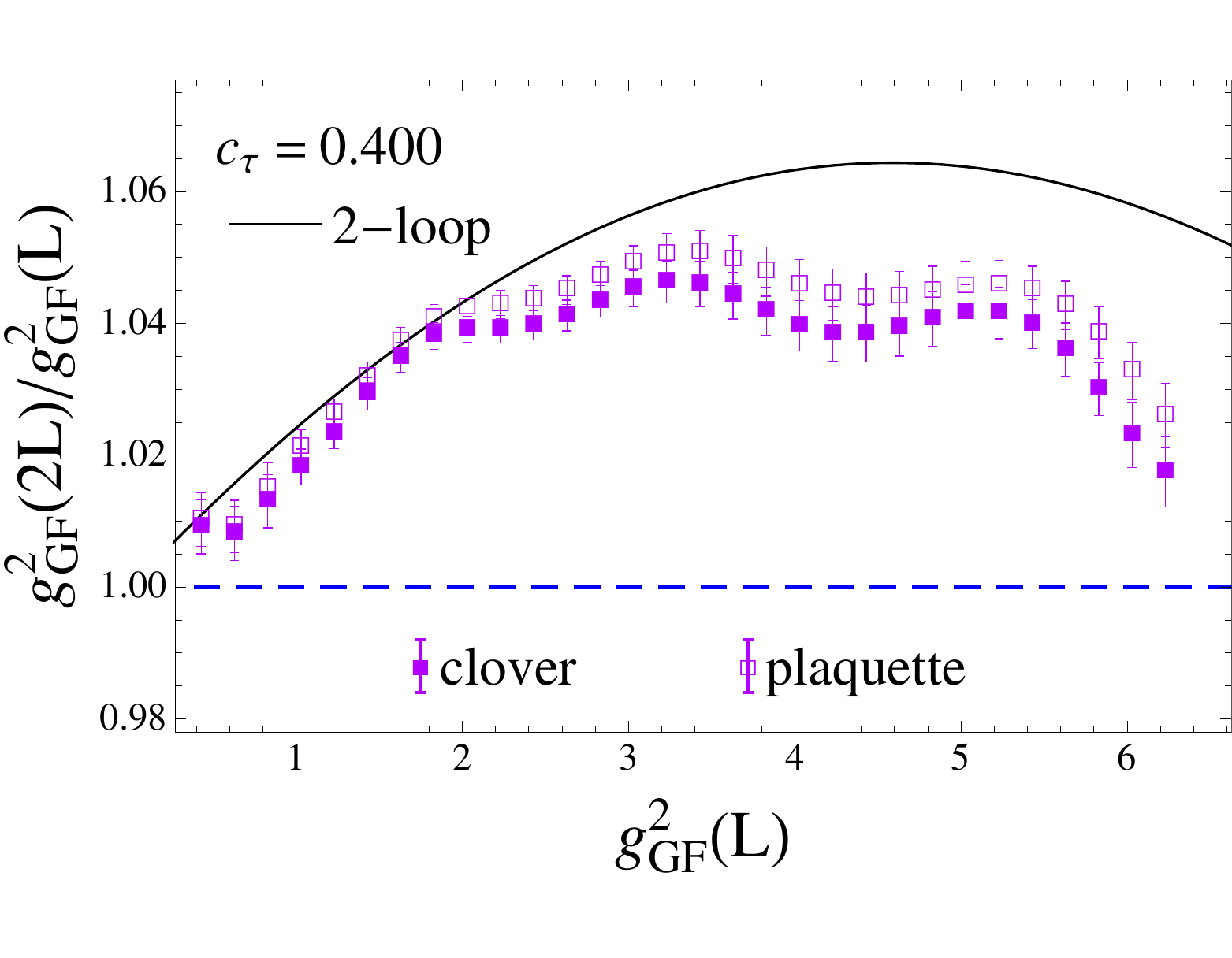}\\
\includegraphics[scale=0.45]{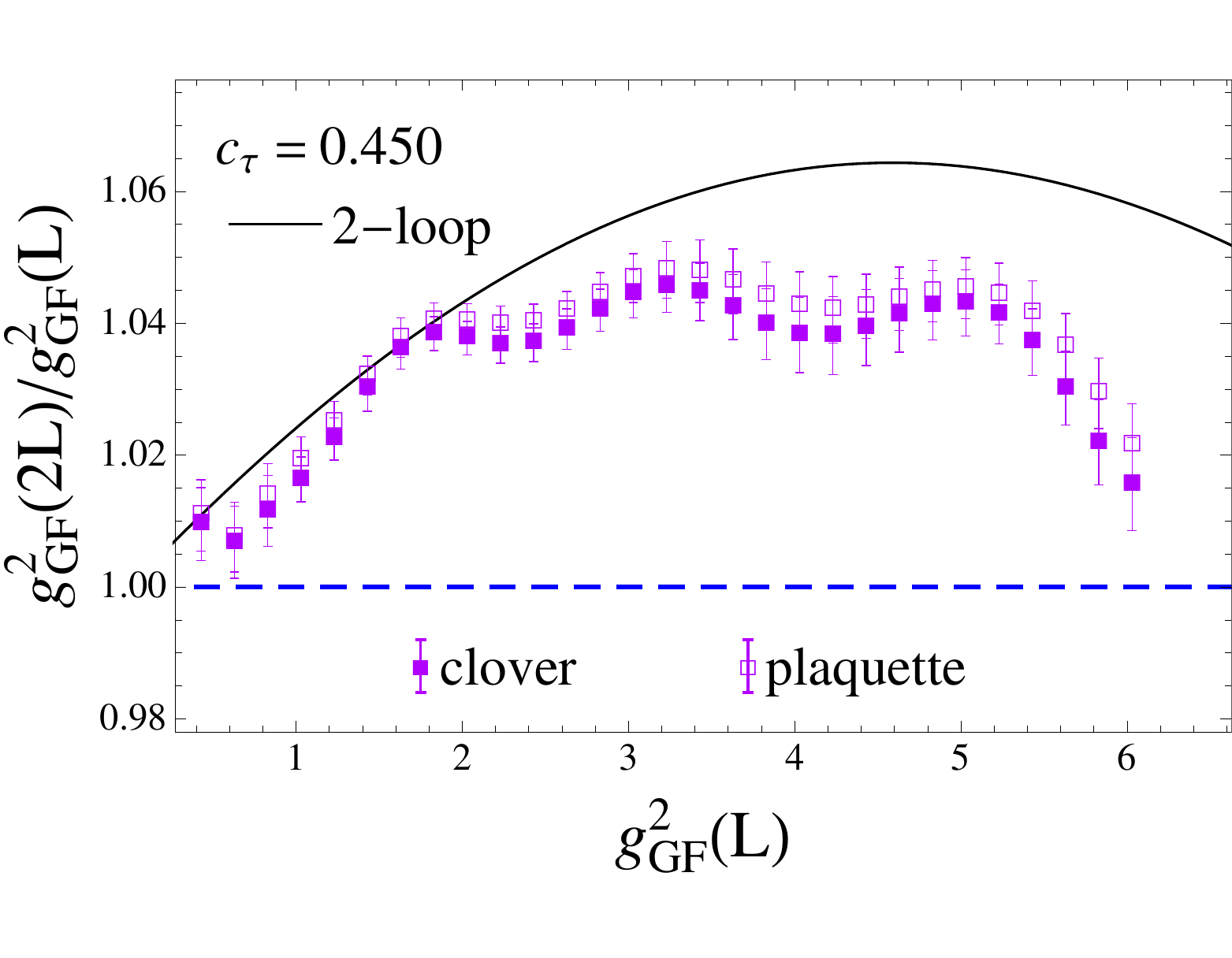}
\hspace{0.5cm}
\includegraphics[scale=0.45]{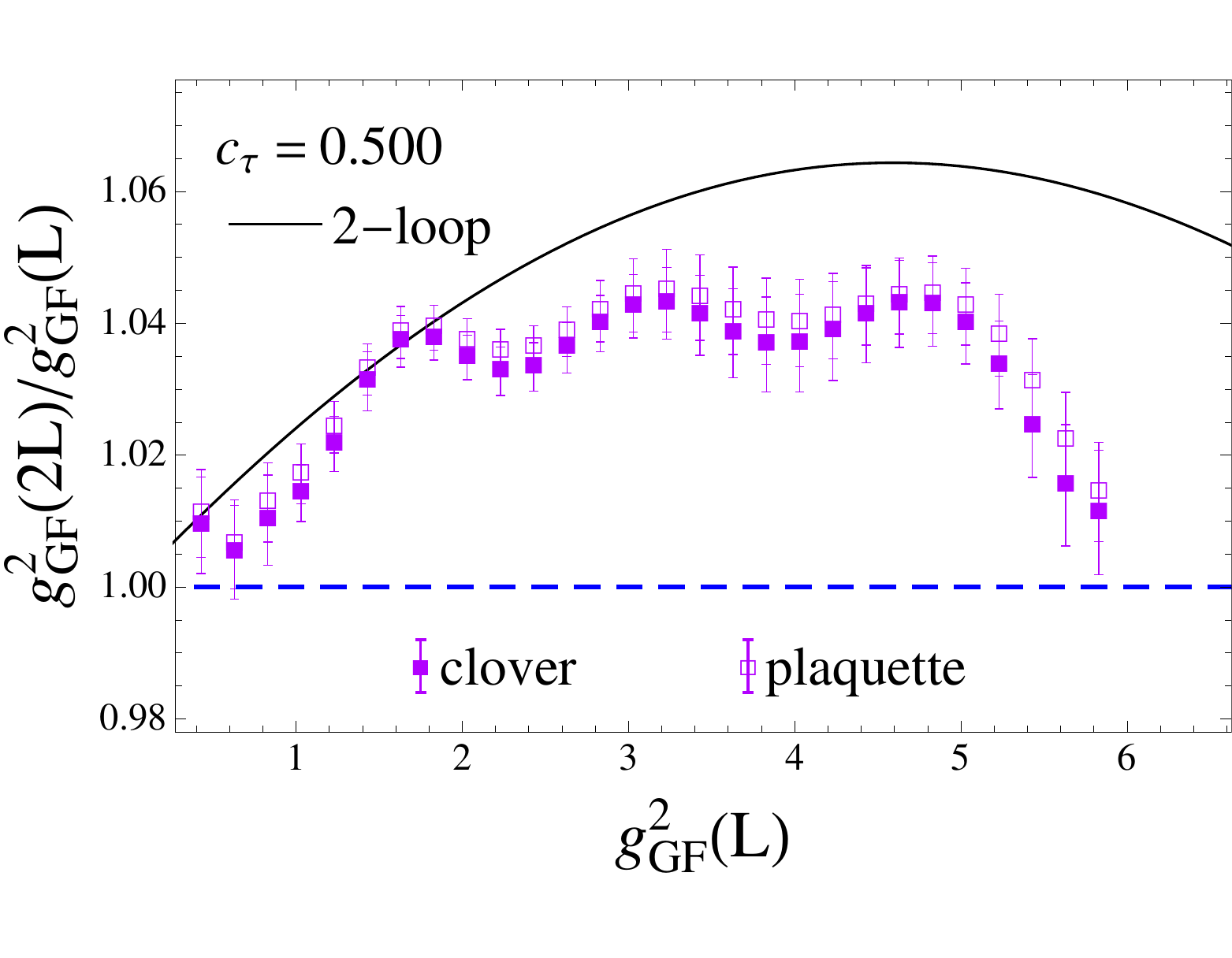}
\caption{The step-scaling functions
  using the procedure discussed in the main text.  
The $x{-}$axis is the value of the input
reference $g_{{\rm GF}}^{2}(L)$.}
\label{fig:sigma}
\end{figure}

Figure~\ref{fig:sigma} shows results of $r_{\sigma} = \sigma (u) /u$,
as defined in Eq.~(\ref{eq:r_sigma_def}), with the renormalised
coupling computed using the clover and the plaquette discretisations.
We first observe that the running of the coupling constant is very slow in
SU(3) gauge theory with 12 flavours.  Doubling the length scale
leads to at most 4 to 5\% change in the coupling constant in the
range that our investigation is performed.   Compared to a ``fast
running'' theory, such as QCD in which the coupling can increase by a
few dozen percent when the length scale is doubled, the running is
very slow in this theory.
In order to provide evidence
for the existence of an IRFP, one has to demonstrate that
$r_{\sigma}$ is unity in both UV and IR regimes, while deviates
from one in between.   Therefore it is
desirable to have high-precision data for such study of this theory.
 In this work we achieve good enough accuracy,
and it is clearly discernible that the theory flows out of the
vicinity of the UV Gaussian fixed point when the length scale is
increased.   At low energy, where $g^{2}_{{\rm GF}} \sim 6$, the results of
$r_{\sigma}$ indicate that the coupling runs significantly slower than
the two-loop perturbative prediction.  
For the case of $c_{\tau} = 0.5$ presented in Fig.~\ref{fig:sigma},
$r_{\sigma}$ is almost consistent with unity in this regime.   This provides evidence
that the scaling of the theory may be governed by IR conformality in
this region.  However, as
already pointed out in Sec.~\ref{sec:continuum_extrap}, the continuum
extrapolation for the results in Fig.~\ref{fig:sigma} is carried out
using Eq.~(\ref{eq:linear_continuum_extrap}) which may not be
applicable near an IRFP.  It requires further scaling test to clarify
this issue.  We will discuss this point in detail in Sec.~\ref{sec:FSS}.

Our analysis relies on the interpolation method reported in
Sec.~\ref{sec:beta_interpolation}, in order to efficiently perform the
time-consuming tuning procedure of Eq.~(\ref{eq:continuum_g_input}).
Although we have many data points for this interpolation
(Table~\ref{tab:bare_g_summary}), it will still introduce correlation
amongst $r_{\sigma}$ computed at different input $g^{2}_{{\rm GF}}$.
That is, values of $g^{2}_{{\rm GF}}(2L)/g^{2}_{{\rm GF}}(L)$ at
different $g^{2}_{{\rm GF}}(L)$ presented in Fig.~\ref{fig:sigma} may
be correlated.  Therefore it is necessary to study the statistical
significance of results in these plots.  Regarding this purpose, we investigate
the likelihood function,
\beq
 L_{h} \left (  r_{\sigma,i}, r_{\sigma,j} \right ) =
 \frac{1}{2\pi \sqrt{{\rm det} ({\rm Cov)}} }
   \mbox{ } 
 {\rm exp} \left [
 - \frac{1}{2} \left ( r_{\sigma,i}  - \bar{r}_{\sigma,i}
 \right ) \left ( {\rm Cov}\right )^{-1}_{ij}  \left ( r_{\sigma,j} - \bar{r}_{\sigma,j}
 \right )  \right ] ,
\label{eq:likelihood_function}
\eeq
where $r_{\sigma,k}$ is the ratio $r_{\sigma}$ at the input coupling
$g^{2}_{{\rm GF}} = u_{k}$,
\beq
 r_{\sigma,k} = r_{\sigma} (u_{k}) ,
\eeq
and $\bar{r}_{\sigma,k}$ is the central value of $r_{\sigma,k}$ in our
numerical computation.  The elements of $2\times 2$ covariant matrix,
${\rm Cov}_{ij}$,  can be determined using the bootstrap samples of
$r_{\sigma,i}$ and $r_{\sigma,j}$ in the numerical calculation.

\begin{figure}[t]
\begin{center}
\scalebox{0.55}{\includegraphics{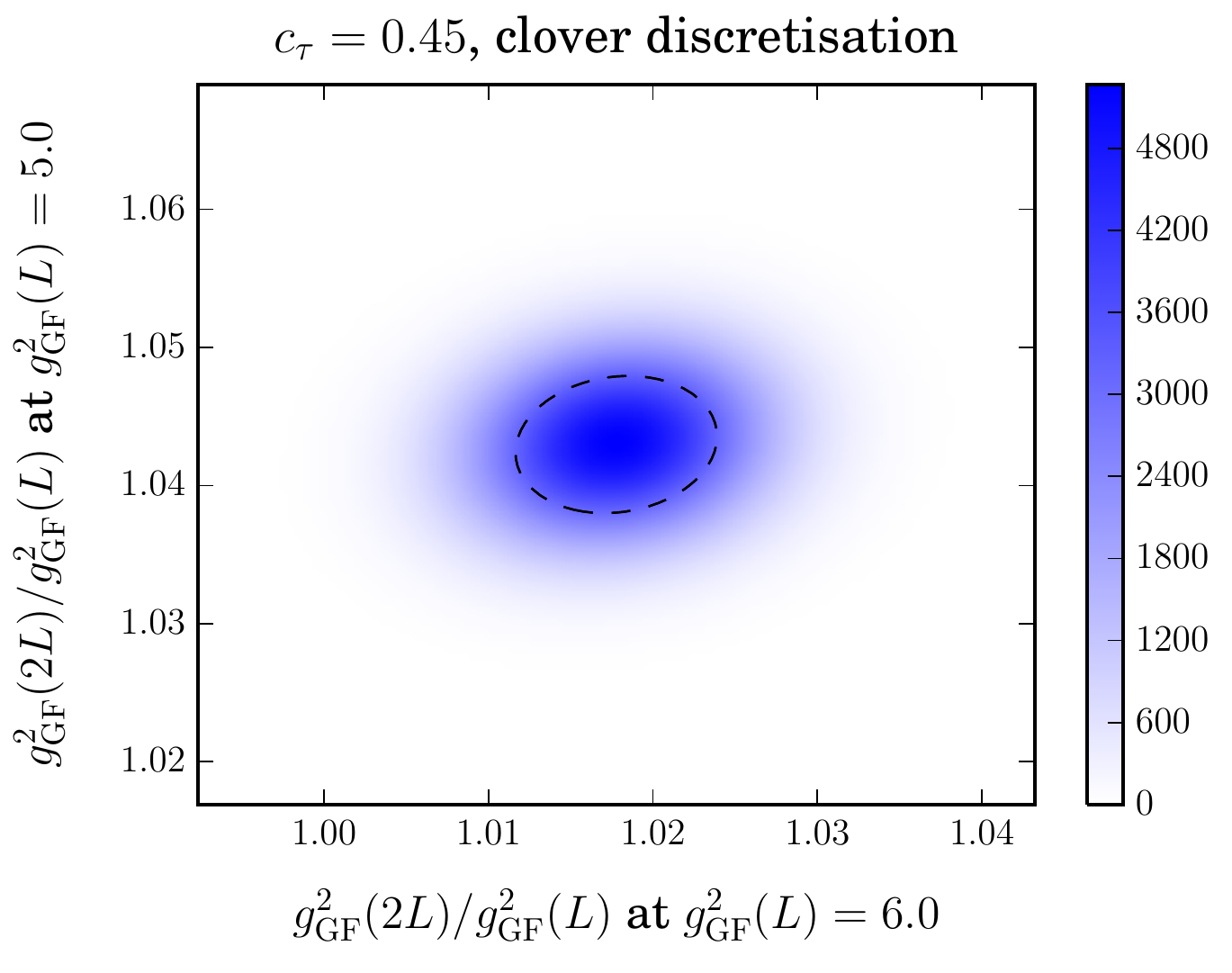}}
\hspace{1cm}
\scalebox{0.55}{\includegraphics{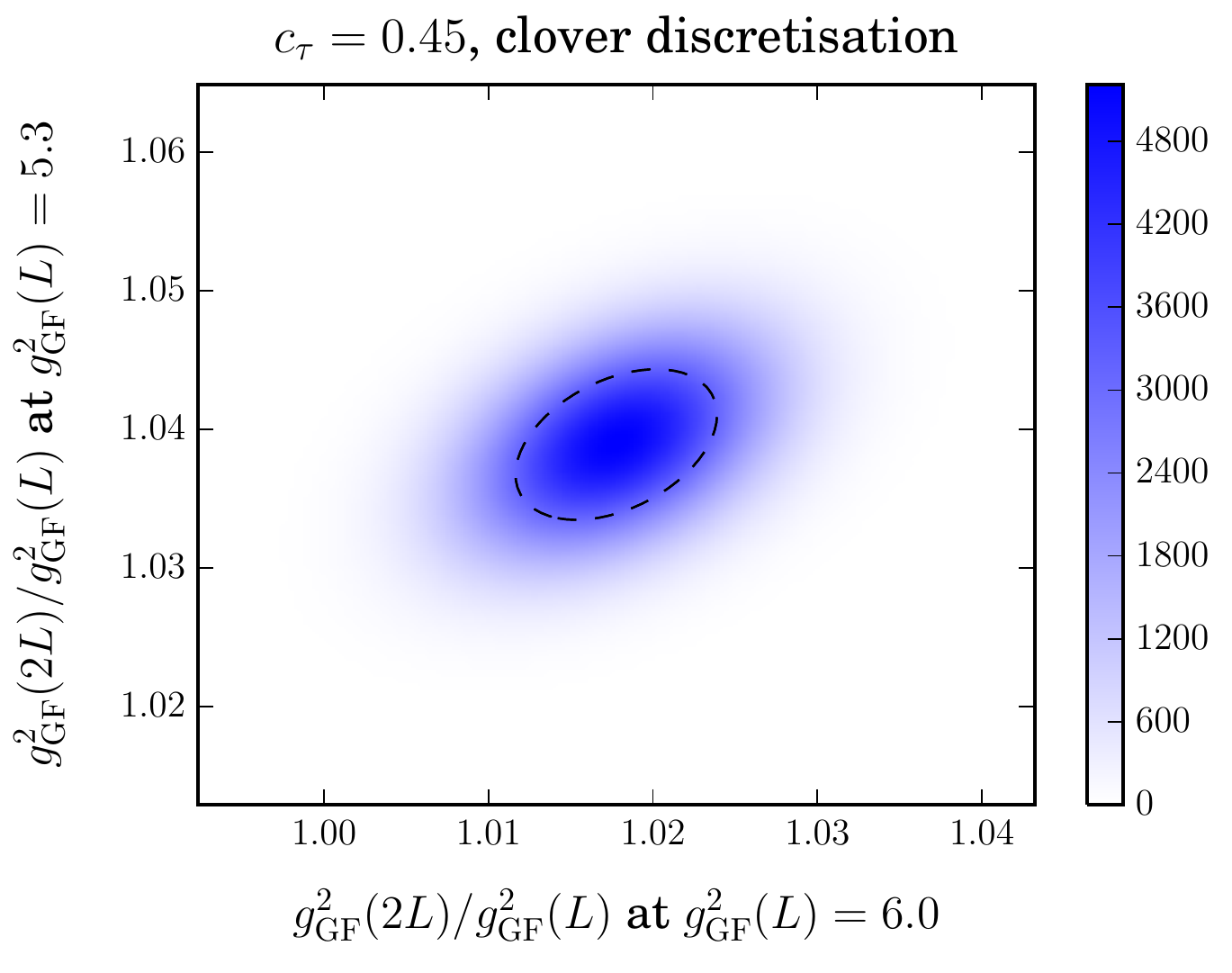}}\\ 
\ \\
\scalebox{0.55}{\includegraphics{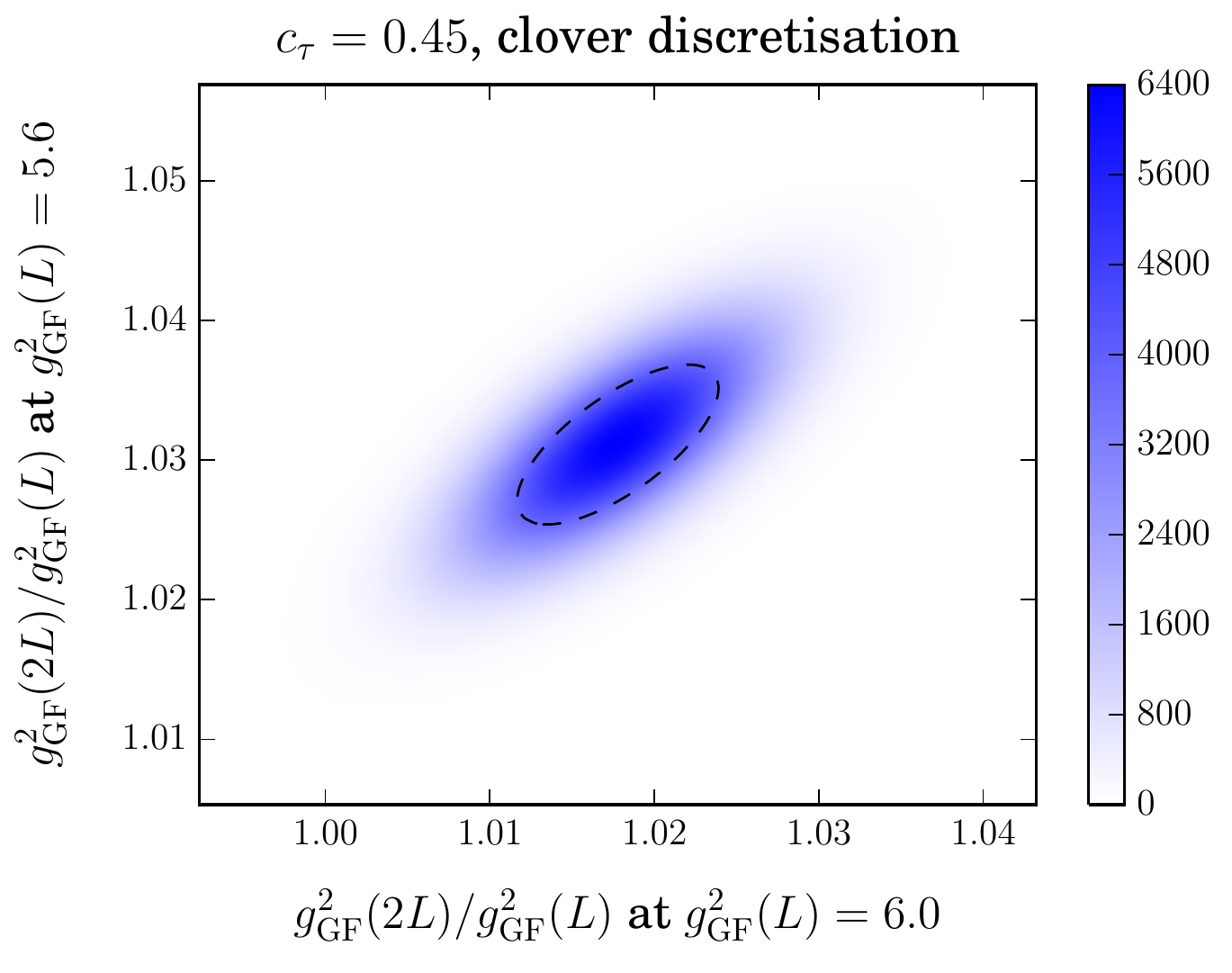}}
\hspace{1cm}
\scalebox{0.55}{\includegraphics{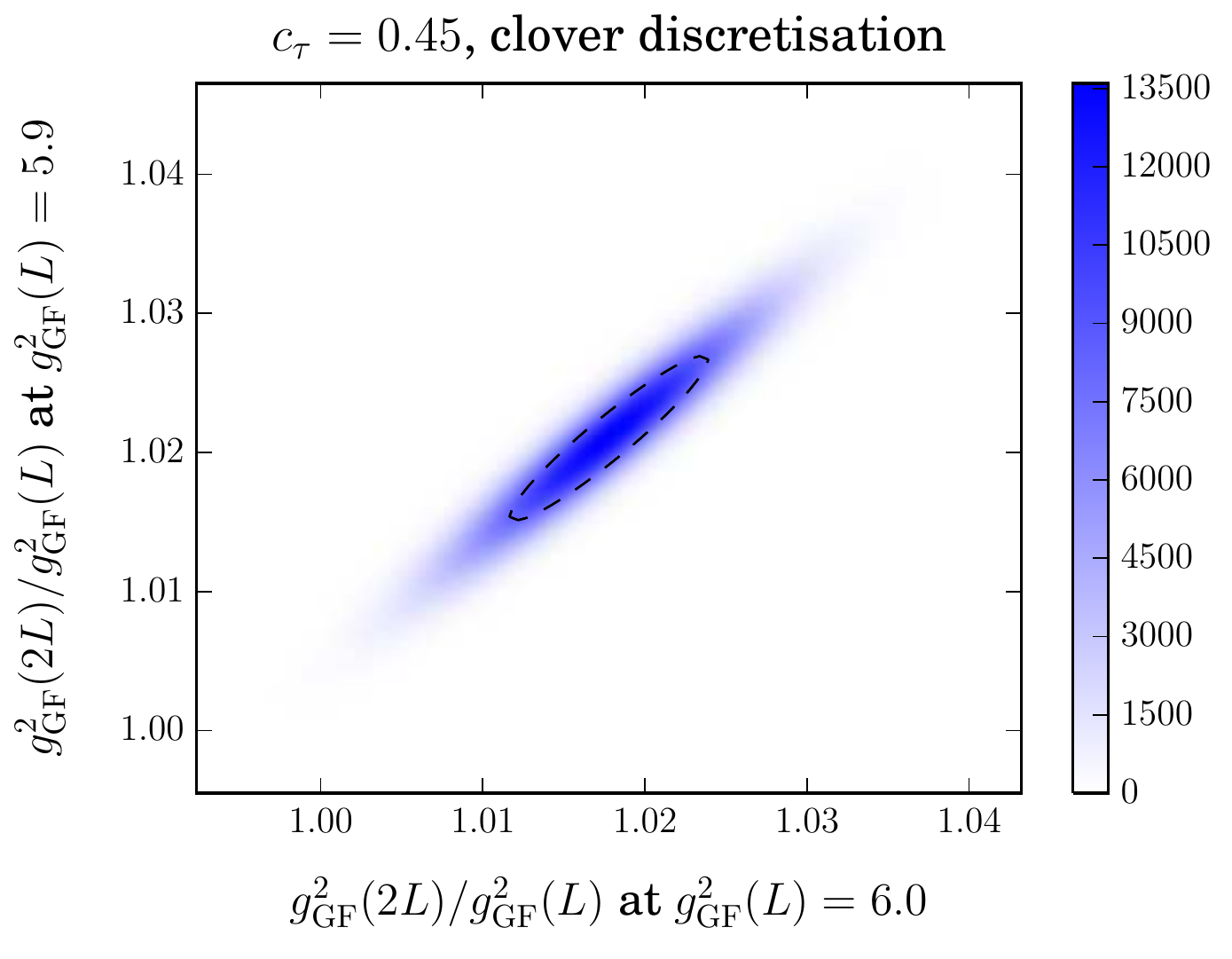}}
\caption{Correlation between $g^{2}_{_{{\rm GF}}}(2L)/g^{2}_{_{{\rm
        GF}}}(L)$ at $g^{2}_{_{{\rm GF}}}(L) = 6.0$ and various other
  results at different input values of $g^{2}_{_{{\rm GF}}}(L)$.
  Plotted here are the results of the likelihood function defined in Eq.~(\ref{eq:likelihood_function}).
The dashed curves represent the
standard error ellipses.}
\label{fig:correlation_of_results}
\end{center}
\end{figure}

Figure~\ref{fig:correlation_of_results} displays the likelihood
functions for the study of the correlation between 
$r_{\sigma}$ at input $g^{2}_{{\rm GF}} = 6.0$ and several choices of
$r_{\sigma,k}$, at $c_{\tau} = 0.45$. 
It is clear that this ratio
computed at input $g^{2}_{{\rm GF}} = 6.0$ is at least mildly
correlated with that extracted at input $g^{2}_{{\rm GF}} \ge 5.3$.
Notice that we have at least two data points for every lattice volume
between these two values of the renormalised coupling.  This investigation
shows the necessity of having simulations at many choices of the bare
coupling for each $\hat{L}$, in order to reduce the correlation
amongst $r_{\sigma}$ computed at different input renormalised
couplings.

\section{Strategy for the continuum extrapolation and
  finite-size scaling}
\label{sec:FSS}
As discussed in Sec.~\ref{sec:continuum_extrap}, implementation of the continuum
extrapolation employing the fit formula of
Eq.~(\ref{eq:linear_continuum_extrap}) is inspired by the
``Symanzik-type'' argument.  This approach is applicable when the bare
parameters are tuned such that the effects
of the lattice spacing are only related to the UV Gaussian fixed
point.  In the present study, this is reached when $g_{0}^{2}$ is
close enough to zero.
Under this circumstance, a major origin of scaling
violation are the irrelevant operators that can be included in the theory.
The classical dimensional analysis is a good approximation in this region, and it leads 
to simple power-law dependence on the cut-off.   For a generic
observable, ${{\mathcal{M}}}_{{\rm latt}}$, computed on the lattice, the
approach to the continuum limit is governed by the behaviour
\beq
 {{\mathcal{M}}}_{{\rm latt}} = {{\mathcal{M}}}_{0} +
 \sum_{n=1}^{\infty} \sum_{i=1}^{N_{{\rm IR}}}
 {{\mathcal{M}}}_{n,i}  \left ( a \Lambda_{i} \right ) ^{n} , 
\label{eq:symanzik_type_cont_limit}
\eeq
where $\Lambda_{i}$ ($i = 1,2,\ldots, N_{{\rm IR}}$) are all the
possible IR energy scales that are well below $1/a$.  Clearly,
${{\mathcal{M}}}_{0}$ is the continuum limit of ${{\mathcal{M}}}_{{\rm
  latt}}$.  Quantum
fluctuations in the above equation can be accounted for by using perturbation theory.  Because
of the Gaussian nature of the fixed point,  they
introduce logarithmic dependence on the lattice spacing in the
coefficients, ${{\mathcal{M}}}_{n,i}$.   These logarithms are often
discernible in numerical analysis only when very high-precision data
are available,
therefore they are normally not included in the fitting procedure.

To employ Eq.~(\ref{eq:symanzik_type_cont_limit}) for conducting the
continuum extrapolation in search of an IRFP using the step-scaling method, it is essential to make certain that the
dimensionfull lattice size, $L$, is in the long-distance region that
is governed by possible
IR conformality, while the scaling property of the theory at the
lattice spacing is still dominated by the UV Gaussian fixed point.  
Therefore, to adopt this Symanzik-type continuum extrapolation for
distinguishing between theories with IR scale invariance
and slow-running behaviour, one may have to perform lattice simulations
extremely close to the limit $\hat{L} \rightarrow \infty$.   This is
particularly crucial for the study of a theory that contains a small
$\beta{-}$function, such that the UV and the IR scaling regimes can be
separated by many orders of magnitude in the difference of scales.
Since one normally works with the lattice size, 
\beq
\label{eq:typical_lattice_size}
 \hat{L} = L/a \sim 10 \mbox{ }{\rm to}\mbox{ } 40,
\eeq
in current step-scaling investigation of the running coupling, it is
challenging to achieve this separation.  Therefore, one has to be
cautious when utilising
the Symanzik-type strategy, Eq.~(\ref{eq:symanzik_type_cont_limit}),
for confirming the existence of an IRFP.

Given the usual choices of the lattice size in
Eq.~(\ref{eq:typical_lattice_size}), it is plausible that if an IRFP
exists in the theory, the bare
couplings can be tuned such that the scaling with respect to the
change in both $a$ and $L$ is controlled by IR conformality.   In
fact, in all the contemporary lattice calculations employing the
step-scaling for probing IR scale invariance in gauge theories~\cite{Appelquist:2007hu,Shamir:2008pb,Appelquist:2009ty,Bursa:2009tj,Hietanen:2009az,Ohki:2010sr,Bursa:2010xn,DeGrand:2010na,Aoyama:2011ry,Giedt:2011kz,Lin:2012iw,DeGrand:2011qd,Hayakawa:2013yfa,Appelquist:2013pqa,Rantaharju:2014ila,Lin:2014fxa,Fodor:2015zna,Hasenfratz:2015ssa,Rantaharju:2015yva}, the
values of $g_{0}^{2}$ are often larger than unity.  Therefore the
continuum extrapolation in these computations (including our present
work) may not be guided by the simple polynomial formula as in 
Eq.~(\ref{eq:symanzik_type_cont_limit}).  Below we examine the
alternative scenario in which the continuum limit is reached according
to approximate IR conformality.

Near an IRFP at strong coupling, the classical dimensional analysis
receives significant corrections from quantum fluctuations, and the
cut-off dependence may no longer be as simple as
Eq.~(\ref{eq:symanzik_type_cont_limit}).  
The anomalous dimensions of the operators in the theory can lead to
dependence on fractional powers of $a/L$.  Investigation for
details of the scaling laws and the continuum limit near possible strong-coupling fixed points is not new in lattice
field theory computations.  Recent examples are the studies of the 
Higgs-Yukawa model in Ref.~\cite{Bulava:2012rb}, and the
three-dimensional scalar theory in Ref.~\cite{Aoki:2014yra}.  Here we
will first illustrate this point in the context of this work by 
examining a generic coupling, $g_{{\rm R}}$, renormalised at the
length-scale $\rho$.  In the vicinity of a
strongly-coupled IRFP, the $\beta{-}$function can be well approximated
by the linearised form,
\beq
 \label{eq:RGE_near_IRFP}
 \beta \left ( g^{2}_{{\rm R}} \right ) \equiv - \rho \frac{ {\rm d} g^{2}_{{\rm R}} }{
   {\rm d} \rho } = \gamma_{\ast} \left ( g^{2}_{{\rm R}} - g^{2}_{\ast} \right ) , 
\eeq
where $g_{\ast}$ is the location of the IRFP, and $\gamma_{\ast}$ is the
slope of the $\beta{-}$function at this zero.  Notice that the value
of $g_{\ast}$ depends on the choice of the renormalisation scheme,
while $\gamma_{\ast}$ is a universal quantity and takes real positive value.   
Integrating Eq.~(\ref{eq:RGE_near_IRFP}) between two length scales,
$l_{1}$ and $l_{2}$, we obtain
\beq
  g^{2}_{{\rm R}} ( l_{2} ) = g^{2}_{\ast} + \left [ g^{2}_{{\rm R}} (
     l_{1}  )  - g^{2}_{\ast}\right ] \left (
   \frac{l_{1}}{l_{2}}\right )^{\gamma_{\ast}} ,
\label{eq:FSS_l1_l2}
\eeq
which clearly indicates the possibility of having dependence on non-integer
powers of $l_{1}$ and $l_{2}$.  For the purpose of our discussion, we introduce another scale,
$L_{{\rm ref}}$, such that
\beq
\label{eq:FSS_scale_hierarchy}
 L > L_{{\rm ref}} > a ,
\eeq
and work with fixed lattice spacing. To proceed,
in the following discussion we will present our argument using the
GF-scheme renormalised coupling, $\bar{g}_{{\rm latt}}^{2}(g_{0}^{2},\hat{L})$, as
defined in Eq.~(\ref{eq:GF_scheme_def}).

Expressing all the length scales in lattice
units, and identifying $l_{1}$ and $l_{2}$  in Eq.~(\ref{eq:FSS_l1_l2}) with
$L_{{\rm ref}}$ and $L$, one obtains
\beq
  \bar{g}^{2}_{{\rm latt}} ( g_{0}^{2}, \hat{L} ) = g^{2}_{\ast} + \left [ \bar{g}^{2}_{{\rm latt}} (
     g_{0}^{2}, \hat{L}_{{\rm ref}} )  - g^{2}_{\ast}\right ] \left (
   \frac{\hat{L}_{{\rm ref}}}{\hat{L}}\right )^{\gamma_{\ast}} ,
\label{eq:FSS_Lref_L_near_IRFP}
\eeq
in the vicinity of the IRFP.  This equation can be regarded as a finite-size
scaling formula.  Confronting it with lattice data enables us to
confirm/exclude IR conformality, and it leads to the determination of $g_{\ast}$ and
$\gamma_{\ast}$.  In addition to fixing the bare coupling, we can
further choose to work at a particular value of $\hat{L}_{\rm ref}$ in
the analysis. 
It has to be stressed again that the renormalised couplings,
$\bar{g}^{2}_{{\rm latt}} ( g_{0}^{2}, \hat{L} )$ 
and $\bar{g}^{2}_{{\rm latt}} ( g_{0}^{2}, \hat{L}_{{\rm ref}} )$,
still contain lattice artefacts, and therefore one has to work in a
regime where lattice artefacts are small compared with the statistical
uncertainties. This can be checked in practice by using different
discretisations and/or different lattice sizes to extract $\bar{g}^{2}_{{\rm latt}}$.  

We further notice, from Fig.~\ref{fig:coupling}, that in this work the change of
$\bar{g}^{2}_{{\rm latt}} (g_{0}^{2},\hat{L})$ is small when varying $\hat{L}$
between 8 and 24 at fixed lattice spacing. When the coupling is very
small, this is due to the effect of the Gaussian UVFP.  
At intermediate and strong couplings, such behaviour arises from the
smallness of the $\beta{-}$function.  Therefore,
away from the
asymptotic-freedom regime, we can fit our data, at a particular choice of $\hat{L}_{\rm
ref}$ and $a$, with the formula, 
\beq
   \bar{g}^{2}_{{\rm latt}} ( g_{0}^{2}, \hat{L} ) = g^{2}_{\rm l} (g_{\rm ref}) +
   \left [ g^{2}_{\rm ref}  - g^{2}_{l} (g_{\rm ref})\right ] \left (
   \frac{\hat{L}_{{\rm ref}}}{\hat{L}}\right )^{\gamma (g_{\rm
   ref})} ,
\label{eq:FSS_Lref_L}
\eeq
where $g_{l}$ and $\gamma$ are the free parameters,
with the definition,
\beq
 g_{{\rm ref}} \equiv \bar{g}_{{\rm latt}} (g_{0}^{2}, \hat{L}_{{\rm ref}}) . 
\label{eq:g_ref_def} 
\eeq
Equation~(\ref{eq:FSS_Lref_L}) can be regarded as the consequence of a
``locally linearised'' $\beta{-}$function, which is a good
approximation only when one works with small variations of the coupling
around $g_{{\rm ref}}$.  This is the reason why $g_{l}$ and $\gamma$
depend on $g_{{\rm ref}}$.  Nevertheless, when the theory is tuned to
be close to an IRFP, this equation must converge to
Eq.~(\ref{eq:FSS_Lref_L_near_IRFP}), and $g_{l}$ and $\gamma$ will
approach constant values, $g_{\ast}$ and $\gamma_{\ast}$.

In the numerical analysis, we always fix $\hat{L}_{\rm ref}$ to be
8, and use data at $\hat{L} = 10, 12, 16, 20, 24$ for
fitting with Eq.~(\ref{eq:FSS_Lref_L})\footnote{We have also tried taking $\hat{L}_{\rm ref} = 10$,
and using data at $\hat{L} = 12, 16, 20, 24$ for fitting.  However,
this leads to a significant increase in the error for the results.}. 
For each fit, we specify a value for $g_{0}^{2}$ (hence $g_{{\rm ref}}$), and
extract $g_{l}$ and $\gamma$.   When conducting this
procedure in a region without IR conformality,  $g_{l}$ and $\gamma$ will show
dependence on the input $g_{\rm ref}$.  On the other hand, when the
theory is engineered to be in the neighbourhood of an IRFP by
tuning the bare coupling, these two
quantities should show a clear trend to converge to $g_{\ast}$ and $\gamma_{\ast}$.
In summary, we can utilise our data and perform the fit to
Eq.~(\ref{eq:FSS_Lref_L}) at fixed $\hat{L}_{\rm ref} = 8$, and scan though many
values of $g_{0}$ (hence $g_{\rm ref}$) in the strong-coupling regime. At each choice of
$g_{0}$, we carry out a fit.
If our data indicate
the existence of an IRFP, both $g_{l}$ and $\gamma$ should
show plateau behaviour when plotted against $g_{\rm ref}$, and
different discretisations for $\bar{g}_{\rm latt}$ will lead to consistent results.

\begin{figure}[t]
\begin{center}
\scalebox{0.65}{\includegraphics{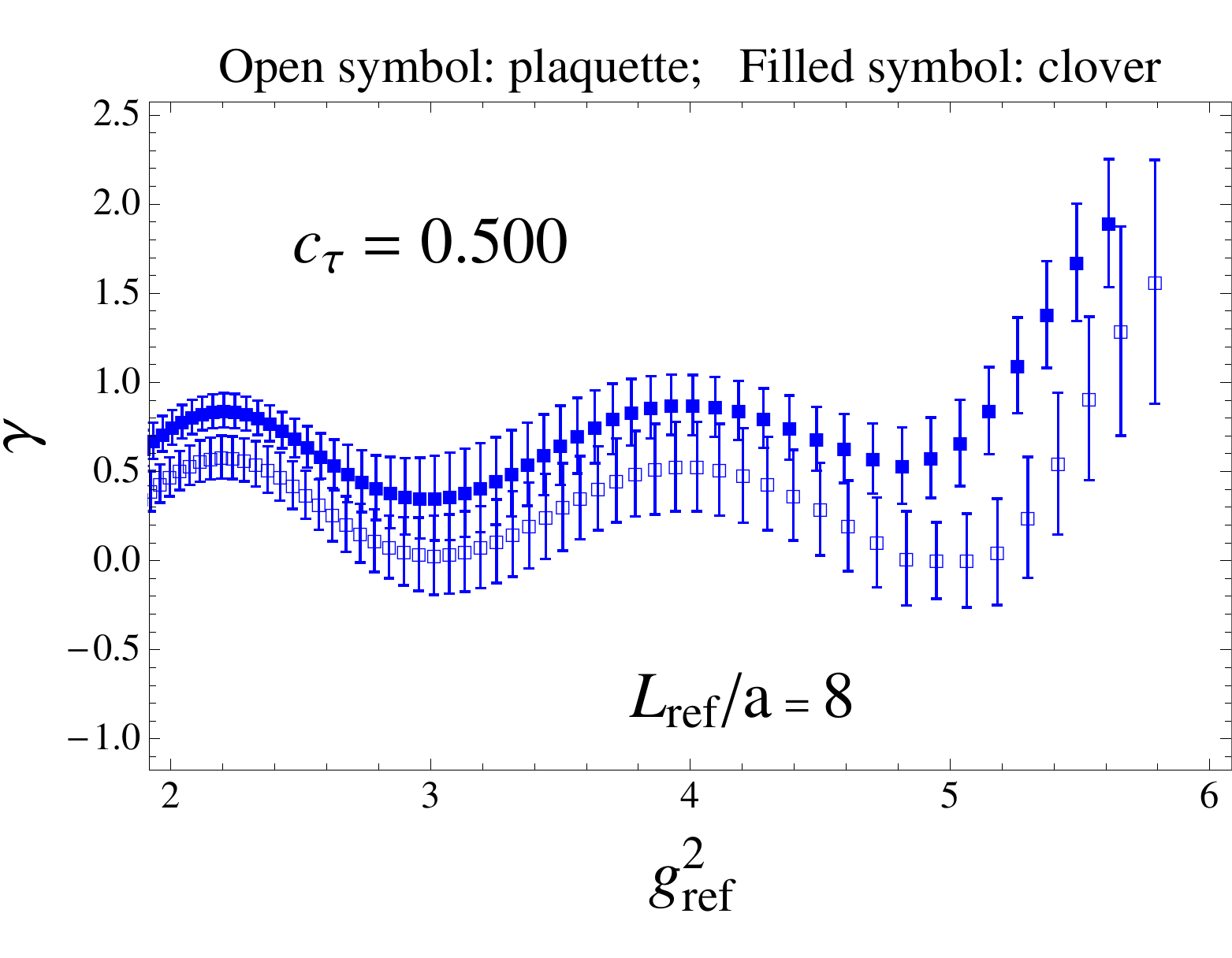}}
\caption{Results of $\gamma$ at $c_{\tau} = 0.5$.  This plot shows
that the theory as probed using our data is insensitive to possible IR conformality.
}
\label{fig:FSS_test}
\end{center}
\end{figure}

Figure~\ref{fig:FSS_test} is the outcome for $\gamma$ determined using
the above analysis procedure at $c_{\tau} = 0.5$. 
In this plot, we notice that the largest $g_{\rm ref}$ for the clover
discretisation is slightly smaller than
the largest input $g_{\rm GF}(L)$ value which is $5.8$ for the same $c_{\tau}$
in the step-scaling analysis presented in
Sec.~\ref{sec:continuum_extrap}.  This is because $\bar{g}_{\rm latt}
(g_{0}^{2}, \hat{L})$ grows with
$\hat{L}$ at fixed $g_{0}^{2}$ as a general trend in our data, and we follow the principle
that no extrapolation in $g_{0}^{2}$ is implemented in this work.  For the finite-size scaling test
discussed in this section, the largest $g_{\rm ref}$ must be chosen to be
the value of $\bar{g}_{\rm latt}
(g_{0}^{2}, \hat{L}=8)$ at the largest $g_{0}^{2}$ where we have data
for the $\hat{L}=24$ lattice.  Therefore, according to
Table~\ref{tab:bare_g_summary}, the maximal bare coupling in
the analysis leading to the result in Fig.~\ref{fig:FSS_test} is 
$g_{0}^{2} = 1.442$.  On the other hand, the largest bare coupling
for computing the input $g_{\rm GF}(L)$ in the step-scaling analysis
is the maximal value of $g_{0}^{2}$ for the $\hat{L}=16$ lattice.  It
can be seen in Table~\ref{tab:bare_g_summary} that this is at
$g_{0}^{2} = 1.449$.  This small difference in $g_{0}^{2}$ can
produce minor but visible difference in the renormalised coupling,
since in this regime $\bar{g}_{\rm latt}$ increases rapidly with
$g_{0}^{2}$, as demonstrated by the plots in Fig.~\ref{fig:coupling}.

From the plot in Fig.~\ref{fig:FSS_test}, it is obvious that results
from the clover and the plaquette 
discretisations are not compatible, and there is no plateau
in the strong-coupling region. This is consistent with our previous
analysis, namely that the theory is not governed by IR conformality
for the values of coupling probed in our simulations.

\section{Comparison with previous works}
\label{sec:comparison}
In this section, we compare our result to previous lattice
step-scaling investigations of SU(3) gauge theory
with twelve flavours.  In Refs.~\cite{Appelquist:2009ty, Lin:2012iw}
this was carried out using the SF and TPL schemes, respectively.
These two calculations made use of the same lattice actions, namely,
the Wilson plaquette gauge action and unimproved staggered fermions.
As summarised in Table~\ref{tab:comparison} ,  while it
was claimed in Refs.~\cite{Appelquist:2009ty,Lin:2012iw} that the
theory can be IR scale-invariant, here we do not see compelling
evidence for this conclusion. 
In comparison with these two previous lattice studies, several aspects
of the computation have been improved in our current project.
First,  the maximal lattice size and the bare
coupling in this work are larger than those in
Refs.~\cite{Appelquist:2009ty,Lin:2012iw}.  In principle, this allows
us to probe the theory at greater length scale.    Secondly, in the
present calculation, the statistical error for the renormalised
coupling is at
the subpercentage level, while it is around or bigger than $2\%$ in
the two earlier works.   Finally, the use of the GF scheme enables us
to obtain our result with two different lattice
discretisations, making it feasible to estimate the systematic effects
arising from the continuum extrapolation.  Such procedure is not
possible in Refs.~\cite{Appelquist:2009ty,Lin:2012iw}, since there are no
alternative discretisations for the observables employed for
determining the renormalised couplings.  In fact, it
was demonstrated that the TPL-scheme coupling computed in the continuum limit in
Ref.~\cite{Lin:2012iw} is unreliable, upon adding lattice data at $\hat{L}=24$
with similar values of the bare coupling~\cite{Ogawa:Latt2013}.  
\begin{table}[t]
\begin{center}
\begin{tabular}{cccc}
\hline\hline
 &$\mbox{ }\mbox{ }\mbox{ }$Ref.~\cite{Appelquist:2009ty} & Ref.~\cite{Lin:2012iw} & This
 work\\
\hline
Scheme & $\mbox{ }\mbox{ }\mbox{ }\mbox{ }\mbox{ }\mbox{ }\mbox{ }\mbox{ }\mbox{ }\mbox{ }\mbox{ }\mbox{ }$SF$\mbox{ }\mbox{ }\mbox{ }\mbox{ }\mbox{ }\mbox{ }\mbox{ }\mbox{ }\mbox{ }$ & $\mbox{ }\mbox{ }\mbox{ }\mbox{ }\mbox{ }\mbox{ }\mbox{ }\mbox{ }\mbox{ }\mbox{ }\mbox{ }\mbox{ }$TPL$\mbox{ }\mbox{ }\mbox{ }\mbox{ }\mbox{ }\mbox{ }\mbox{ }\mbox{ }\mbox{ }\mbox{ }\mbox{ }\mbox{ }$ & $\mbox{ }\mbox{ }\mbox{ }\mbox{ }\mbox{ }\mbox{ }\mbox{ }\mbox{ }\mbox{ }$GF$\mbox{ }\mbox{ }\mbox{ }\mbox{ }\mbox{ }\mbox{ }\mbox{ }\mbox{ }\mbox{ }$\\
Largest $L/a$ & $\mbox{ }\mbox{ }\mbox{ }$20 & 20 & 24\\
Largest $g_{0}^{2}$ & $\mbox{ }\mbox{ }\mbox{ }\sim 1.40$ & $\sim 1.05$ & $\sim 1.45$\\
\hline
Conclusion & $\mbox{ }\mbox{ }\mbox{ }$IRFP at $g^{2}_{\rm SF} \sim 5$ & IRFP at $g^{2}_{\rm TPL} \sim
2$ & No IRFP up to $g^{2}_{\rm GF} \sim 6$\\
\hline\hline
\end{tabular} 
\caption{Comparison of the result from this work with two
previous lattice step-scaling investigations of SU(3) gauge theory
with twelve flavours, using the same
actions but with the Schr\"{o}dinger-Functional (SF) and the
Twisted-Polyakov-Loop (TPL) schemes.  The symbols
$g^{2}_{\rm SF}$ and $g^{2}_{\rm TPL}$ denote the renormalised
couplings extracted in the Sch\"{o}dinger Functional and the Twisted
Polyakov Loop schemes.}
\label{tab:comparison}
\end{center}
\end{table}

It has to be stressed that in Table~\ref{tab:comparison}, the results
from Refs.~\cite{Appelquist:2009ty,Lin:2012iw}
are both obtained using the Symanzik-type ans\"{a}tz for the
continumm extrapolation.   As was already pointed out in
Sec.~\ref{sec:continuum_extrap}, and discussed in detail in
Sec.~\ref{sec:FSS}, one has to be cautious when using this
approach to confirm
the existence of an IRFP with lattice simulations.

We notice that the authors of Ref.~\cite{Cheng:2014jba}
have performed a lattice computation for the GF-scheme
coupling for twelve-flavour SU(3) gauge theory, 
employing a procedure that is similar
to the step-scaling method.   In Ref.~\cite{Cheng:2014jba}, the fundamental-adjoint
plaquette gauge action and the nHYP-smeared~\cite{Hasenfratz:2001hp,Hasenfratz:2007rf} staggered fermions are used.
For the analysis procedure, the clover discretisation has been adopted to 
extract $\bar{g}^{2}_{\rm latt} (g_{0}^{2},\hat{L})$, as defined in
Eq.~(\ref{eq:GF_scheme_def}) in the current paper, with the
normalisation factor, $\hat{{\mathcal{N}}}$, calculated using the continuum
perturbation theory.   In the strong-coupling regime, these authors search for the intersections of
pairs of curves representing $\bar{g}^{2}_{\rm latt}(g^{2}_{0},\hat{L})$ as a function of
$g^{2}_{0}$, at $\hat{L}$ and $s \hat{L}$ with $s$
being the step size.  Such intersections are
interpreted as the consequence of IR scale invariance.  The values of $\bar{g}^{2}_{\rm latt}$
at these intersections are then extrapolated to the limit of vanishing
$a/L$ with the Symanzik-type ans\"{a}tz, and the result is regarded as
the location of the IRFP in the continuum limit.   In
Ref.~\cite{Cheng:2014jba}, it is claimed that this
IRFP is reached at $g^{2}_{\rm GF} \sim 7$.  This
procedure cannot be implemented in the present work, because the above
pairwise intersections are not 
observed in our data for $c_{\tau} \ge 0.45$, where we have the 
lattice artefacts under control.

The conclusion in Ref.~\cite{Cheng:2014jba} does not contradict the
result of this work.   Our main observation is that the scaling
behaviour of the GF-scheme coupling in SU(3) gauge theory
with twelve flavours is not governed by IR conformality at $g^{2}_{\rm
GF} \sim 6$.   Of course the effects of the possible IRFP can appear
at $g^{2}_{\rm GF} > 6$.  This is beyond the scope of this
project.  On the other hand, it will be interesting to examine the
reliability of the 
procedure in Ref.~\cite{Cheng:2014jba} by carrying out the same
analysis using the plaquette discretisation.

\section{Conclusion}
\label{sec:conclusion}
In this article, we present our step-scaling analysis of the 
coupling constant in SU(3) gauge theory
with 12 massless flavours, using the Gradient-Flow scheme.
In this theory the
$\beta{-}$function is very small, such that doubling the length scale
induces at most $6\%$ variation in the renormalised coupling according
to two-loop perturbation theory.  Therefore, to make any
statistically-meaningful statement regarding possible IR conformality
in this theory, it is desirable to have lattice data with error at the
subpercentage level for the extracted renormalised coupling.   
To our knowledge, our work is the first
computation that achieves such precision.

It is well known that the continuum extrapolation is the main source
of the systematic error in the step-scaling approach.  The
implementation of the Yang-Mills gradient flow reduces the cut-off
effects in our calculation.  In this project, we obtain the
renormalised coupling {\it via} the computation of the gauge field
energy density using two different lattice discretisations,
namely the plaquette and the clover operators.  Such strategy enables
the estimate of the systematic error arising from lattice artefacts.
We find that at large enough flow time, such that
\beq
 c_{\tau} \ge 0.45 ,
\eeq
this extrapolation is under control.

Being able to have good control of both statistical and systematic
errors, we manage to
demonstrate, in a statistically-meaningful manner, that the theory flows out of the
vicinity of the UV Gaussian fixed point at $g_{{\rm GF}}^{2} \sim 2$,
and the running of the coupling begins to be significantly slower than
the two-loop perturbative prediction around $g_{{\rm GF}}^{2} \sim 5$.
At $c_{\tau} = 0.5$, our result indicates that the ratio $r_{\sigma}
= g_{{\rm GF}}^{2}(2L)/g_{{\rm GF}}^{2}(L)$ is almost consistent with
unity at $g_{{\rm GF}}^{2}(L) \sim 5.8$.  This conclusion is reached
using the Symanzik-type formula in performing the continuum extrapolation.

In this paper, we discuss the application of the continuum
extrapolation ans\"{a}tz {\it a'la} Symanzik in the search for
possible IRFP through the step-scaling approach.  
This Symanzik-type method is based on the scenario that the property of the theory
at the cut-off scale is governed by the UV Gaussian fixed point, while its scaling
behaviour at the lattice size can be dominated by IR conformality.  
This is obviously very challenging to achieve in practice.  To confirm
the existence of the IRFP, we argue that it is
essential to examine the theory with the
assumption that the scaling of the theory at the lattice spacing is
also determined by IR scale invariance.  Following this
argument, we perform a finite-size scaling test of SU(3) gauge theory
with twelve flavours.  The result of this test indicates that the
behaviour of the theory, as probed using our lattice data, is
not governed by possible IR conformality.  That is, our result does not
support the existence of an IRFP in this theory in the region $g_{{\rm GF}}^{2}(L) \le 6$.

In summary, lattice computations for the determination of the
conformal windows for various gauge theories have matured
significantly in recent years.  The importance of controlling errors
in these calculations is receiving growing attention.  The work
presented in this paper is our first attempt in this research avenue
with high-accuracy lattice data.  To make further progress in this direction,
it would be desirable to implement the procedure with improved actions
in future lattice simulations.

\appendix
\section{Raw data at $c_{\tau} = 0.5$}
\label{sec:raw_data}
\begin{table}[h]
\begin{tabular}{lll}\hline
\multicolumn{3}{c}{$L/a = 8$} \\
\hline
$\mbox{ }\mbox{ }\beta \mbox{ }\mbox{ }\mbox{ }\mbox{ }\mbox{ }\mbox{ }$ &  $\left ( \bar{g}_{{\rm latt}}^{2} \right )^{{\rm clover}}\mbox{ }\mbox{ }$ & $\left ( \bar{g}_{{\rm latt}}^{2} \right )^{{\rm plaq}}\mbox{ }\mbox{ }$ \\
\hline
4.11  &  5.926(25)  &  6.233(25)   \\

4.22  &  5.331(18)  &  5.599(19)   \\

4.28  &  5.119(16)  &  5.369(16)   \\

4.36  &  4.777(13)  &  5.007(13)   \\

4.50  &  4.277(22)  &  4.475(22)   \\

4.70  &  3.757(17)  &  3.922(17)   \\

5.00  &  3.208(21)  &  3.338(20)   \\

5.36  &  2.693(8)  &  2.796(8)   \\

5.50  &  2.552(27)  &  2.646(27)   \\

5.53  &  2.509(5)  &  2.602(5)   \\

5.81  &  2.237(4)  &  2.317(4)   \\

6.00  &  2.077(13)  &  2.150(13)   \\

6.12  &  2.003(3)  &  2.072(3)   \\

6.47  &  1.800(2)  &  1.859(2)   \\

6.50  &  1.808(18)  &  1.866(18)   \\

6.76  &  1.660(4)  &  1.712(4)   \\

7.00  &  1.559(8)  &  1.607(8)   \\

7.11  &  1.512(2)  &  1.558(2)   \\

7.82  &  1.280(3)  &  1.317(3)   \\

8.00  &  1.233(10)  &  1.268(10)   \\

8.45  &  1.132(1)  &  1.164(1)   \\

9.00  &  1.028(6)  &  1.056(6)   \\

9.42  &  0.950(2)  &  0.976(2)   \\

10.00  &  0.867(10)  &  0.890(9)   \\

11.15  &  0.752(4)  &  0.770(4)   \\

12.00  &  0.675(4)  &  0.692(4)   \\

13.85  &  0.557(2)  &  0.570(2)   \\

14.00  &  0.556(3)  &  0.569(3)   \\

15.23  &  0.495(2)  &  0.507(2)   \\

16.00  &  0.464(3)  &  0.475(3)   \\

17.55  &  0.414(1)  &  0.424(1)   \\

18.00  &  0.402(3)  &  0.412(3)   \\

20.00  &  0.353(1)  &  0.362(1)   \\

20.13  &  0.351(1)  &  0.359(1)   \\
\hline
\end{tabular}
\hspace{1cm}
\begin{tabular}{lll}\hline
\multicolumn{3}{c}{$L/a = 16$} \\
\hline
$\mbox{ }\mbox{ }\beta \mbox{ }\mbox{ }\mbox{ }\mbox{ }\mbox{ }\mbox{ }$ &  $\left ( \bar{g}_{{\rm latt}}^{2} \right )^{{\rm clover}}\mbox{ }\mbox{ }$ & $\left ( \bar{g}_{{\rm latt}}^{2} \right )^{{\rm plaq}}\mbox{ }\mbox{ }$ \\
\hline
4.14  &  6.043(22)  &  6.291(22)   \\

4.26  &  5.473(19)  &  5.697(19)   \\

4.38  &  5.011(20)  &  5.214(20)   \\

4.48  &  4.677(16)  &  4.864(16)   \\

4.60  &  4.333(16)  &  4.506(17)   \\

4.70  &  4.057(12)  &  4.218(12)   \\

5.00  &  3.438(10)  &  3.572(11)   \\

5.30  &  2.965(19)  &  3.078(19)   \\

5.36  &  2.878(10)  &  2.988(10)   \\

5.50  &  2.696(8)  &  2.799(8)   \\

5.53  &  2.675(10)  &  2.777(10)   \\

5.70  &  2.494(9)  &  2.589(9)   \\

5.81  &  2.393(9)  &  2.483(9)   \\

6.12  &  2.137(8)  &  2.217(8)   \\

6.47  &  1.907(5)  &  1.978(5)   \\

6.76  &  1.751(5)  &  1.816(5)   \\

7.11  &  1.592(4)  &  1.650(5)   \\

7.82  &  1.348(5)  &  1.397(5)   \\

8.00  &  1.297(11)  &  1.344(11)   \\

8.45  &  1.179(3)  &  1.222(3)   \\

9.00  &  1.069(3)  &  1.098(4)   \\

9.42  &  0.992(3)  &  1.027(3)   \\

11.15  &  0.775(3)  &  0.803(3)   \\

12.00  &  0.698(6)  &  0.723(6)   \\

13.85  &  0.572(2)  &  0.592(2)   \\

15.23  &  0.506(2)  &  0.524(2)   \\

16.00  &  0.477(4)  &  0.494(5)   \\

17.55  &  0.421(2)  &  0.436(1)   \\

18.00  &  0.407(4)  &  0.421(4)   \\

20.00  &  0.358(1)  &  0.371(1)   \\

20.13  &  0.358(1)  &  0.370(1)   \\
\hline
 & & \\
 & & \\
 & & \\
\end{tabular}
\label{tab:raw_data_L8_16}
\caption{Raw data for the renormalised couplings extracted using the clover and the plaquette discretisations at $\hat{L} = 8$ and 16 with $c_{\tau} = 0.5$.}
\end{table}
\begin{table}[h]
\begin{tabular}{lll}\hline
\multicolumn{3}{c}{$L/a = 10$} \\
\hline
$\mbox{ }\mbox{ }\beta \mbox{ }\mbox{ }\mbox{ }\mbox{ }\mbox{ }\mbox{ }$ &  $\left ( \bar{g}_{{\rm latt}}^{2} \right )^{{\rm clover}}\mbox{ }\mbox{ }$ & $\left ( \bar{g}_{{\rm latt}}^{2} \right )^{{\rm plaq}}\mbox{ }\mbox{ }$ \\
\hline
4.12  &  5.984(24)  &  6.262(25)   \\

4.22  &  5.493(25)  &  5.741(25)   \\

4.33  &  5.029(23)  &  5.253(24)   \\

4.50  &  4.417(23)  &  4.608(24)   \\

4.70  &  3.870(21)  &  4.033(21)   \\

5.00  &  3.296(28)  &  3.427(29)   \\

5.36  &  2.764(5)  &  2.869(5)   \\

5.50  &  2.573(22)  &  2.671(22)   \\

5.53  &  2.571(6)  &  2.668(6)   \\

5.81  &  2.302(5)  &  2.386(6)   \\

6.00  &  2.142(11)  &  2.219(11)   \\

6.12  &  2.052(5)  &  2.126(5)   \\

6.47  &  1.839(4)  &  1.904(4)   \\

6.50  &  1.844(18)  &  1.908(19)   \\

6.76  &  1.686(5)  &  1.633(22)   \\

7.00  &  1.606(11)  &  1.660(11)   \\

7.11  &  1.542(4)  &  1.593(4)   \\

7.82  &  1.303(2)  &  1.346(2)   \\

8.00  &  1.268(14)  &  1.309(14)   \\

8.45  &  1.154(2)  &  1.191(2)   \\

9.00  &  1.041(11)  &  1.074(11)   \\

9.42  &  0.968(2)  &  0.998(2)   \\

10.00  &  0.888(2)  &  0.915(2)   \\

11.15  &  0.763(3)  &  0.786(3)   \\

12.00  &  0.690(8)  &  0.711(8)   \\

13.85  &  0.561(2)  &  0.578(2)   \\

14.00  &  0.542(7)  &  0.559(7)   \\

15.23  &  0.497(2)  &  0.512(2)   \\

16.00  &  0.469(4)  &  0.483(4)   \\

17.55  &  0.418(1)  &  0.430(1)   \\

18.00  &  0.406(4)  &  0.418(4)   \\

20.00  &  0.357(1)  &  0.367(1)   \\

20.13  &  0.357(1)  &  0.367(1)   \\
\hline
\end{tabular}
\hspace{1cm}
\begin{tabular}{lll}\hline
\multicolumn{3}{c}{$L/a = 20$} \\
\hline
$\mbox{ }\mbox{ }\beta \mbox{ }\mbox{ }\mbox{ }\mbox{ }\mbox{ }\mbox{ }$ &  $\left ( \bar{g}_{{\rm latt}}^{2} \right )^{{\rm clover}}\mbox{ }\mbox{ }$ & $\left ( \bar{g}_{{\rm latt}}^{2} \right )^{{\rm plaq}}\mbox{ }\mbox{ }$ \\
\hline
4.15  &  6.105(24)  &  6.350(24)   \\

4.28  &  5.434(23)  &  5.651(23)   \\

4.41  &  5.030(22)  &  5.230(23)   \\

4.53  &  4.589(16)  &  4.770(16)   \\

4.66  &  4.218(16)  &  4.384(16)   \\

4.80  &  3.898(15)  &  4.050(15)   \\

5.10  &  3.350(10)  &  3.479(11)   \\

5.40  &  2.858(8)  &  2.968(9)   \\

5.70  &  2.537(4)  &  2.634(4)   \\

6.00  &  2.266(7)  &  2.352(7)   \\

6.50  &  1.931(6)  &  2.004(6)   \\

7.00  &  1.655(5)  &  1.717(5)   \\

8.00  &  1.308(5)  &  1.356(5)   \\

9.00  &  1.082(4)  &  1.122(4)   \\

10.00  &  0.915(3)  &  0.948(3)   \\

12.00  &  0.699(3)  &  0.725(3)   \\

14.00  &  0.567(4)  &  0.588(4)   \\

16.00  &  0.478(2)  &  0.496(3)   \\

18.00  &  0.411(2)  &  0.426(2)   \\

20.00  &  0.361(2)  &  0.374(2)   \\

50.00  &  0.129(1)  &  0.133(1)   \\
\hline
 & & \\
 & & \\
 & & \\
 & & \\
 & & \\
 & & \\
 & & \\
 & & \\
 & & \\
 & & \\
 & & \\
 & & \\
\end{tabular}
\label{tab:raw_data_L10_20}
\caption{Raw data for the renormalised couplings extracted using the clover and the plaquette discretisations at $\hat{L} = 10$ and 20 with $c_{\tau} = 0.5$.}
\end{table}
\begin{table}[h]
\begin{tabular}{lll}\hline
\multicolumn{3}{c}{$L/a = 12$} \\
\hline
$\mbox{ }\mbox{ }\beta \mbox{ }\mbox{ }\mbox{ }\mbox{ }\mbox{ }\mbox{ }$ &  $\left ( \bar{g}_{{\rm latt}}^{2} \right )^{{\rm clover}}\mbox{ }\mbox{ }$ & $\left ( \bar{g}_{{\rm latt}}^{2} \right )^{{\rm plaq}}\mbox{ }\mbox{ }$ \\
\hline
4.10  &  6.241(26)  &  6.513(27)   \\

4.20  &  5.654(24)  &  5.898(25)   \\

4.33  &  5.064(19)  &  5.280(20)   \\

4.50  &  4.498(15)  &  4.685(16)   \\

4.70  &  3.961(13)  &  4.121(13)   \\

5.00  &  3.337(12)  &  3.469(12)   \\

5.30  &  2.902(10)  &  3.013(10)   \\

5.36  &  2.815(7)  &  2.923(7)   \\

5.50  &  2.631(17)  &  2.731(18)   \\

5.53  &  2.615(7)  &  2.714(7)   \\

5.81  &  2.332(6)  &  2.418(6)   \\

6.00  &  2.181(17)  &  2.262(18)   \\

6.12  &  2.094(5)  &  2.171(5)   \\

6.47  &  1.868(4)  &  1.935(4)   \\

6.50  &  1.878(15)  &  1.945(16)   \\

6.76  &  1.715(4)  &  1.775(4)   \\

7.00  &  1.592(13)  &  1.648(14)   \\

7.11  &  1.558(4)  &  1.613(4)   \\

7.82  &  1.319(4)  &  1.365(4)   \\

8.00  &  1.248(11)  &  1.291(12)   \\

8.45  &  1.166(3)  &  1.205(3)   \\

9.00  &  1.046(11)  &  1.082(12)   \\

9.42  &  0.982(3)  &  1.015(3)   \\

10.00  &  0.926(11)  &  0.956(11)   \\

11.15  &  0.763(3)  &  0.788(3)   \\

12.00  &  0.687(8)  &  0.709(8)   \\

13.85  &  0.568(3)  &  0.587(3)   \\

14.00  &  0.566(6)  &  0.585(6)   \\

15.23  &  0.506(3)  &  0.522(3)   \\

16.00  &  0.469(6)  &  0.484(6)   \\

17.55  &  0.420(2)  &  0.434(2)   \\

18.00  &  0.404(5)  &  0.417(5)   \\

20.00  &  0.364(7)  &  0.375(7)   \\

20.13  &  0.353(2)  &  0.364(2)   \\
\hline
\end{tabular}
\hspace{1cm}
\begin{tabular}{lll}\hline
\multicolumn{3}{c}{$L/a = 24$} \\
\hline
$\mbox{ }\mbox{ }\beta \mbox{ }\mbox{ }\mbox{ }\mbox{ }\mbox{ }\mbox{ }$ &  $\left ( \bar{g}_{{\rm latt}}^{2} \right )^{{\rm clover}}\mbox{ }\mbox{ }$ & $\left ( \bar{g}_{{\rm latt}}^{2} \right )^{{\rm plaq}}\mbox{ }\mbox{ }$ \\
\hline
4.16  &  5.998(25)  &  6.236(26)   \\

4.30  &  5.414(21)  &  5.628(22)   \\

4.44  &  4.964(23)  &  5.158(23)   \\

4.57  &  4.512(20)  &  4.688(20)   \\

4.70  &  4.183(18)  &  4.346(19)   \\

4.85  &  3.866(14)  &  4.016(15)   \\

5.20  &  3.172(12)  &  3.295(12)   \\

5.60  &  2.657(9)  &  2.759(9)   \\

6.00  &  2.288(3)  &  2.375(3)   \\

6.50  &  1.934(4)  &  2.007(4)   \\

7.00  &  1.678(4)  &  1.741(4)   \\

7.50  &  1.471(4)  &  1.526(4)   \\

8.00  &  1.321(3)  &  1.370(3)   \\

9.00  &  1.081(3)  &  1.121(3)   \\

10.00  &  0.917(3)  &  0.951(3)   \\

12.00  &  0.704(2)  &  0.730(2)   \\

14.00  &  0.569(2)  &  0.590(2)   \\

16.00  &  0.473(2)  &  0.491(2)   \\

18.00  &  0.412(1)  &  0.427(2)   \\

20.00  &  0.366(2)  &  0.380(2)   \\
\hline
 & & \\
 & & \\
 & & \\
 & & \\
 & & \\
 & & \\
 & & \\
 & & \\
 & & \\
 & & \\
 & & \\
 & & \\
 & & \\
 & & \\
\end{tabular}
\label{tab:raw_data_L12_24}
\caption{Raw data for the renormalised couplings extracted using the clover and the plaquette discretisations at $\hat{L} = 12$ and 24 with $c_{\tau} = 0.5$.}
\end{table}

\clearpage

\section*{Acknowledgments}
We warmly thank Luigi Del Debbio, Issaku Kanamori and Ben Svetitsky
for useful discussion, Hiroshi Ohki and Eigo Shintani for reading the draft
carefully.   Help in developping the HMC simulation code
from Tatsumi Aoyama and Hideo
Matsufuru is crucial for making this work
possible.  A portion of gauge field configurations for $\hat{L} \le 16$ lattices were
generated and shared with us by colleagues involved in the work of Ref.~\cite{Aoyama:2011ry}.
Support from Taiwanese MOST with grant 102-2112-M-009-002-MY3 is 
acknowledged.  We are grateful to Taiwanese NCHC where most of the
computation work was carried out.  C.-J.D.L. thanks the travel support
and the hospitality from CERN during the progress of this project.



\bibliographystyle{apsrev.bst}
\bibliography{refs} 

\end{document}